\documentclass[aps,10pt,pra,twocolumn,groupedaddress, floatfix, superscriptaddress,showpacs, showkeys,amsfonts]{revtex4-2}
\usepackage{amsfonts}
\usepackage{amsmath}
\usepackage{amssymb, setspace}
\usepackage{graphics,graphicx}
\graphicspath{{./}}
\usepackage{epsf}
\usepackage{subfigure}
\usepackage[%
  colorlinks=true,
  urlcolor=blue,
  linkcolor=blue,
  citecolor=blue
]{hyperref}
\usepackage[all]{hypcap}
\usepackage{color}
\usepackage[utf8]{inputenc}
\usepackage{dsfont}
\usepackage{enumitem}
\usepackage[utf8]{inputenc}

\usepackage{comment}
\usepackage{orcidlink}
\usepackage{soul} 

\usepackage{natbib}
\usepackage{braket}

\newcommand{\be}{\begin{equation}}
\newcommand{\ee}{\end{equation}}
\newcommand{\bea}{\begin{eqnarray}}
\newcommand{\eea}{\end{eqnarray}}
\newcommand{\eq}[1]{Eq.~\eqref{#1}}

\newcommand{\eqss}[2]{Eqs.~\eqref{#1}-\eqref{#2}}
\newcommand{\seq}[1]{Sec.~\ref{#1}}

\newcommand{\app}[1]{App.~\ref{#1}}
\newcommand{\fig}[1]{Fig.~\ref{#1}}

\newcommand{\bem}{\begin{multline}}
\newcommand{\eem}{\end{multline}}

\newcommand{\dket}[1]{\ket{#1}\!\rangle}

\newcommand\identity{1\kern-0.25em\text{l}}

\newcommand{\apporsm}[1]{App.~\ref{#1}}

\begin{document}
\title{
Modeling error correction with Lindblad dynamics and approximate channels
}

\newcommand{\bina}{Faculty of Engineering and the Institute of Nanotechnology and Advanced Materials, Bar Ilan University, Ramat Gan, Israel}
\newcommand{\huji}{School of Computer Science and Engineering, Hebrew University, Jerusalem, Israel}
\newcommand{\ibm}{IBM Quantum, IBM Research - Israel, Haifa University Campus, Mount Carmel, Haifa 31905, Israel}

\author{Zohar Schwartzman-Nowik\orcidlink{0000-0003-3804-4933}}
\email{zohar.nowik@mail.huji.ac.il}
\affiliation{\huji}
\affiliation{\bina}
\affiliation{\ibm}

\author{Liran Shirizly\orcidlink{0009-0002-4597-0126}}
\email{liran.shirizly@ibm.com}
\affiliation{\ibm}

\author{Haggai Landa}
\email{haggai.landa@ibm.com}
\affiliation{\ibm}

\begin{abstract}

We analyze the performance of a quantum error correction code subject to physically motivated noise modeled by a Lindblad master equation. We consider dissipative and coherent single-qubit terms and two-qubit crosstalk, studying how different approximations of the noise capture the success rate of a code. Focusing on the five-qubit code and adapting it to partially correct two-qubit errors in relevant parameter regimes according to the noise model, we find that a composite-channel approximation where every noise term is considered separately captures the behavior in many physical cases up to long timescales, eventually failing due to the effect of noncommuting terms. In contrast, we find that single-qubit approximations do not properly capture the error correction dynamics with two-qubit noise, even for short times. A Pauli approximation going beyond a single-qubit channel is sensitive to the details of the noise, state, and decoder, and succeeds at short timescales relative to the noise strength, beyond which it fails. Furthermore, we point out a mechanism for a Pauli model failure where it underestimates the failure rate of a code even with frequent syndrome projection and correction cycles. 
These results shed light on the performance of error correction in the presence of realistic noise and can advance the ongoing efforts towards useful quantum error correction. 
\end{abstract}

\maketitle

\section{Introduction}\label{sec:intro}
Quantum computers represent a new paradigm in computing,
holding great potential for applications in fields of science,
technology, and industry. 
However, practical, large-scale quantum computers
would require overcoming open challenges related to quantum error correction \cite{sundaresan2023demonstrating, PhysRevX.11.041058,krinner2022realizing, postler2022demonstration,google2023suppressing, gupta2023encoding, bluvstein2023logical,bravyi2024high}. 
Error correction involves encoding the information into logical states belonging to a subspace of a higher-dimensional space, 
harnessing redundancy to facilitate error detection and the restoration of the
original state. 
With stabilizer error correction \cite{RevModPhys.87.307}, logical
states are characterized by the measurement outcomes of a predefined
set of observables, referred to as stabilizers since they do not alter the logical states. A result of the stabilizers measurement is called a syndrome
and when it is different from the one associated with
a logical state, an error is identified. The syndrome serves as input to a decoder, which identifies the recovery process
that will transform the corrupted state back into a logical state.
 
When the recovery process restores the original state, this is a success, and when it results in a different logical state, it is a failure.
When analyzing the success rate, a common
simplifying assumption is to model every single operation element of the noise as a Pauli operator (possibly multiqubit, though often single-qubit Pauli noise is assumed). 
The analysis proceeds by constructing a parity matrix, which conveys information about how the measurement outcomes of stabilizers are affected by specific Pauli errors. For any instance of Pauli noise (meaning a specification of the Pauli errors that occurred), the syndrome is obtained by multiplication of the parity matrix with a vector representing the error. This makes the first step, that of obtaining the syndrome, an efficient process that does not require simulating the logical state itself. As a result, this approach has been used extensively to study the behavior of error correction codes under various scenarios. These include, in particular, biased noise models \cite{bonilla2021xzzx, dua10clifford, demarti2022performance, roffe2023bias}, correlated noise \cite{nickerson2019analysing}, and explorations of advantages of codes adapted (or tailored) to particular noise types \cite{PhysRevApplied.8.064004, tiurev2023correcting}.

However, Pauli noise channels are just a subset of completely positive and trace preserving (CPTP) quantum operations (or Kraus channels).
Moreover, multiqubit noise channels can rarely be derived explicitly in physically-relevant cases due the large number of parameters required for the representation. In contrast, the underlying generators of continuous dynamics of individual single qubits (1Q) and interacting two-qubit (2Q) pairs are often simpler to handle and characterize using well-established methods \cite{krantz2019quantum,van2023probabilistic}. Using the generators, the evolution of open systems is described by a master equation, and when the environment is memoryless (the dynamics being Markovian), the Lindblad master equation is the standard choice since it guarantees CPTP dynamics by construction \cite{gorini_completely_1976, lindblad_generators_1976,naeij2024open}. Lindbladian systems are extensively studied, with analytic, numerical and approximate solution methods continuously being improved \cite{shirizly2024dissipative, BP07, lidar2019lecture, 10.1063/1.5115323, PhysRevA.98.063815, PhysRevLett.128.033602,PhysRevLett.131.190403,LM23, houdayer2023solvable,  PhysRevA.108.062219,gravina2023adaptive, chen2023periodically, guimarães2024optimized}. Nevertheless, only a few existing works have compared error correction success rates between physically-motivated noise descriptions and their Pauli approximations. 
The majority of previous work considered only 1Q noise terms \cite{gutierrez_comparison_2015, gutierrez2016errors, tomita2014low,greenbaum2017modeling, PhysRevLett.119.040502}.
Other works have focused on the effect of coherent errors (often considering analytically solvable cases), as in \cite{sanders2015bounding, PhysRevA.95.062338, bravyi2018correcting, PhysRevA.99.022313, beale2018quantum, iverson2020coherence}, and crosstalk and gate errors have been simulated using Kraus channels with small codes \cite{debroy2020logical,google2023suppressing}, or twirled to obtain Pauli channels as in \cite{heussen2023strategies}.
Some works devoted to linking a basic Lindblad model of a system of qubits to error correction success rates include \cite{PhysRevA.89.022306, PhysRevA.105.022612}, a comparison to experiment in \cite{andersen2020repeated}, and numerics going beyond Lindblad dynamics (departing from the Born-Markov approximation for 1Q noise) in \cite{PhysRevResearch.5.043161}.

In this paper, 
we study the accuracy of
different approximations in modelling the behavior of a quantum error correction code when
considering hardware-relevant Lindblad noise. We consider dynamics generated by 1Q coherent (unitary) and incoherent (dissipative) noise and also 2Q coherent interactions. We focus on the success rate of the five-qubit
code \cite{gottesman2010introduction}, a toy-model code capable of correcting both bit- and phase-flip errors \cite{PhysRevA.99.022313, ryan2022implementing}, which we adapt to decode and correct 2Q errors (at the expense of some of the 1Q errors), according to the noise model. We study the question of quantum memory decoherence, with a logical code state evolving subject to the noise for varying durations. We compare the performance of the code with the state evolving under several approximations of the noise: (i) continuous Lindblad dynamics, the closest approach to realistic dynamics among those considered here (and the most computationally demanding one), 
(ii) an intermediate approach
consisting of a composition of partial 2Q and 1Q Kraus channels,
(iii) a Pauli approximation of the channels, and 
(iv) an attempt at an approximation using 1Q channels only.

We perform our study within
the framework of a code-capacity noise \cite{landahl_fault-tolerant_2011}, which considers the logical state initialization, the syndrome measurements and the error recovery processes as ideal. This framework is often used (together with Pauli models) to calculate the error correction threshold of notable code families \cite{tuckett2018ultrahigh}. We also follow the dynamics of syndrome correction cycles, where we observe a nontrivial evolution even within the code-capacity framework. 
This simplified model allows investigating those dynamics and the validity of some common noise-approximation assumptions, which can be hard to isolate in more complex setups such as in the circuit-level model. 
For a realistic example, we take parameter values from an experimental characterization of currently deployed IBM Quantum superconducting-qubit devices accessible via the cloud \cite{IBMQuantum}, using the open-source packages \textit{Qiskit} \cite{Qiskit} and \textit{Qiskit Experiments} \cite{kanazawa2023qiskit}.
The 1Q and 2Q noise terms that we study are relevant in other types of quantum hardware \cite{heussen2023strategies}, as discussed also later.

The main results of this paper are as follows. The composite-channel approximation, which is based on noise generators but does not take into account the non-commutativity between noise terms, is a good approximation for relevant ranges of noise parameters, and up to considerably long time scales. We point out the regime of its failure due to the neglect of noncommuting error terms. On the other hand, the validity of the multi-qubit Pauli approximation strongly depends on the noise parameters, initial state and decoder. For many cases it holds pretty well for short time scales, which are most relevant for error correction, and diverges significantly for longer times, and our study exemplifies a number of caveats of the Pauli approximation, whose relevance is broad and extends beyond the code capacity model. In particular, the Pauli noise model could fail in capturing effects of coherent noise and accumulation of phases in logical errors. In \seq{Sec:LogicalDM} we further show this failure when frequently repeated syndrome cycles are applied. 

The structure of the paper is as follows. In \seq{sec:lindbladian}, we introduce the Lindbladian noise model that forms the basis of
our investigation.
In \seq{sec:analysis_lind}, we outline the methods required
to deduce the syndrome evolution under the noise. In \seq{sec:approx_hier} we develop an approximation
hierarchy for the code states evolution.
In \seq{sec:approx_results}, we
present our results regarding this approximation hierarchy and also introduce
 modified decoders tailored to the noise model.
Finally,
in \seq{sec:closing}, we summarize our results and discuss what has been learnt, and what are possibly important open questions for further exploration. The Appendix contains further details of the setups studied here and the approaches employed.

\section{The Lindbladian noise model}\label{sec:lindbladian}

As the fundamental model for the qubits' dynamics we consider the master equation describing the continuous-time evolution of a system's density matrix $\rho(t)$ when being subject to the noisy channel for time $t$,
\begin{equation}
\frac{\partial\rho}{\partial t}=-\frac{i}{\hbar}\left[\mathcal{H},\rho\right]+\mathcal{D}_{0}\left[ \rho \right] +\mathcal{D}_{2}\left[ \rho \right].\label{eq:rho}
\end{equation}
In \eq{eq:rho}, $\mathcal{H}$ is the Hamiltonian accounting for the unitary dynamics, $\left[A,B\right]=AB-BA$ is the commutator and $\mathcal{D}_0$ and $\mathcal{D}_{2}$ are the dissipator terms describing the nonunitary dynamics. Ideally, for qubits left idle, we would have $\mathcal{H}=0$. Therefore any nontrivial $\mathcal{H}$ describes an effective (unitary) noise. We consider $\mathcal{H}$ in the form
\begin{equation}
\mathcal{H}/\hbar =\sum_{i}H_{i}+\sum_{\left\langle i,j\right\rangle }V^{zz}_{ij},\label{eq:hamiltonian}
\end{equation} 
with
\begin{equation}\label{Eq:1Q_ham}
H_{i}=\frac{1}{2}h_{i} \left(I_i-\sigma^z_{i}\right),\qquad
V^{zz}_{ij}=\frac{1}{2}\zeta_{ij}  \sigma^z_{i} \sigma^z_{j},
\end{equation}
 where $\sigma^a_i$ is the Pauli matrix acting on qubit $i$ with $a\in\{x,y,z\}$, $I_i$ is the single-qubit identity matrix, the sum $\sum_{i}$ runs over all the qubits, $\sum_{\left\langle ij\right\rangle }$
runs over all coupled pairs, and $h_{i}$, $\zeta_{i,j}$ are coefficients in units of angular frequency. 
$H_{i}$ corresponds to a discrepancy
between the actual qubit frequency and our estimated value of it (determined by the precision of the calibration experiment and ensuing drifts), and $V^{zz}_{ij}$ is a residual crosstalk between coupled qubits, resulting from uncontrolled interactions of ZZ type (also referred to as Ising coupling).

In physical systems the crosstalk would often couple nearest-neighbor qubits that would typically be data and parity (auxiliary) qubits, and parity qubits are lacking in our model to begin with. Nevertheless, effective longer-range interaction is also possible, and we consider this coupling as a representative term of coherent dynamics between qubits that are coupled either directly or indirectly. Some 2Q coherent dynamics is also expected when taking into account the over/under rotations of realistic 2Q gates. We note that another prevalent interaction term is the XY (``exchange'' or ``flip-flip'') coupling \cite{krantz2019quantum}, which is nondiagonal in the Z basis and introduces some further complexity into the dynamics. For completeness we briefly analyze XY coupling in \app{sec:XY}, strengthening the main conclusions of this work.

The dissipators are given by
\begin{align}
\mathcal{D}_{0}\left[ \rho \right] & =\sum_i g_{0,i}\left(\sigma^+_{i}\rho\sigma^-_{i}-\frac{1}{2}\left\{ \sigma^-_{i}\sigma^+_{i},\rho\right\} \right),\label{eq:amplitude damping lindbladian}\\
\mathcal{D}_{2}\left[ \rho \right] & =\sum_i g_{2,i}\left(\sigma^z_{i}\rho\sigma^z_{i}-\rho\right),\label{eq:phase damping lindbladian}
\end{align}
where 
$\sigma^{\pm}_{i}=\frac{1}{2}\left(\sigma^x_{i}\pm i\sigma^y_{i}\right)$, 
$\left\{ A,B\right\} =AB+BA$ is the anti-commutator, and $g_{0,i}$ and $g_{2,i}$ are non-negative
coefficients determining the rates of the nonunitary terms.
$\mathcal{D}_{0}$ is the generator of amplitude damping (separately for each qubit)
and similarly, $\mathcal{D}_{2}$ generates the phase damping noise channel. We are following the common notation in the field, where the ground state of an isolated qubit is the state obeying $\ket{0}=\sigma^z\ket{0}$, and therefore $\mathcal{D}_{0}$ drives the qubits to the ground state. The environmental-induced heating term is often much smaller in strength and we neglect it here.

In the model as written above, the parameters of the qubits could be different for different qubits in the device. These possible variations across the device and in time would strongly depend on the system.
For example, trapped ions could have mostly identical parameters (and often stable in time), while superconducting-qubit parameters would often vary between different qubits and fluctuate in time between experiment realizations. In \app{sec:experiment} we consider experimental results for the distribution of those parameters on two state-of-the-art superconducting devices with $\gtrsim 100$ qubits, and in \app{sec:inhomogeneous} we analyze the dynamics of an inhomogeneous system in the measured regime of parameters. The results of the appendix plausibly show that the spatial inhomogeneity does not invalidate conclusions derived when considering a system with homogeneous qubit parameters, at least for the code and parameters that we consider here. This observation is consistent with Ref.~\cite{carroll2024subsystem}, where the mean logical error rate depends on the average of the physical qubit infidelity distribution for a broad parameter range.

Hence, in the following sections we consider for simplicity the qubit-independent values
\be h_i\equiv h, \quad \zeta_{ij}\equiv \zeta, \quad g_{0,i} \equiv g_0, \quad g_{2,i} \equiv g_2.\ee 
It is
also customary to define
\be \label{Eq:t1t2_def} T_1=\frac{1}{g_0}, \qquad T_{\phi}= \frac{1}{2g_2},\ee
and $T_2$ that is given by \be 
\frac{1}{T_{2}}=\frac{1}{2T_{1}}+\frac{1}{T_{\phi}}.\label{Eq:t2_def} \ee
$T_1$ is the timescale of amplitude damping, $T_{\phi}$ is the ``pure'' dephasing time, and $T_2$ the
phase damping (or decoherence) time.

Finally, we must also specify the connectivity between qubits, i.e., the pairs of qubits for which the interaction terms $V^{zz}_{ij}$ are nonzero. 
Since we are analyzing the scenario of code capacity we are not modelling auxiliary qubits explicitly, and therefore our simulations contain only the data qubits. 
Except where specified differently, we consider in the following a connectivity of all-to-all among the five code qubits.

\section{Analysis of code performance with Lindbladian noise}\label{sec:analysis_lind}

\subsection{Estimating error correction success rates}

As discussed in \seq{sec:intro}, when the noise is modeled as a Pauli noise, there is an efficient way to analyze the performance of the code, without the need of simulating a logical state. 
In contrast, when we are interested in a more complex physically-motivated noise described using continuous-time generators, there seems to be no way to analyze the code performance in a manner that is more efficient than simulating the logical state, which is how we proceed \footnote{
A representation of the full channel acting on all qubits, as was done in \cite{jain2023improved}, could be even harder to achieve numerically in the general case.}.
The simulation begins by selecting an initial
logical state and initializing the qubits into this
state. We focus on the eigenstates of the logical Pauli operators, i.e.,
the states 
\begin{equation}
\left\{ |0\rangle_{L},|1\rangle_{L},|+\rangle_{L},|-\rangle_{L},|+i\rangle_{L},|-i\rangle_{L}\right\}. \label{eq:states to average}
\end{equation}
Subsequently, the noise is applied as dictated by the Lindbladian
for a time duration $t$ corresponding to the time scale of the pertinent
noisy operator, resulting in the mixed state $\rho(t)$. 

The next question is how to extract a useful metric from $\rho(t)$. As discussed further below, an explicit analysis that accounts for all possible syndromes and their corresponding recoveries could become challenging with large codes due to the exponential scaling of their number. Instead, we present in this subsection a simple and efficient approach to compute
the code performance, suitable within the code capacity model and somewhat advantageous with larger codes.

Every error correction code
has a set of correctable errors that the code can rectify, $C\equiv\left\{ C_{j}\right\} $. These errors
can be chosen in such a way that their application to a logical code
state results in orthogonal states, as discussed in \cite{gottesman2010introduction}.
Starting with an initial ideal logical state $|\psi_{0}\rangle$,
we define $|\varphi_{j}\rangle=C_{j}|\psi_{0}\rangle$. Following the noise operation (before the syndrome extraction and
recovery process) the components that overlap with the states $|\varphi_{j}\rangle$ will be successfully corrected by the error correction procedure.
The components of the state that do not overlap with $|\varphi_{j}\rangle$ are outside of the space of correctable errors, and successful error correction is not guaranteed.
Therefore,
for a given state $\rho$, the probability of guaranteed success
is given by 
\begin{equation}
p_{C}=\sum_{\left\{\varphi_{j}\right\}} \left\langle \varphi_{j}| \rho|\varphi_{j}\right\rangle =\sum_{C} \bra{ \psi_{0}}C_{j}^{\dagger}\rho C_{j}\ket{\psi_{0}}. 
\label{pc}
\end{equation}
\eq{pc} allows computing the success rate without simulating the syndrome measurement and recovery process, but rather only simulate the noisy evolution of the initial logical state.
A similar formula has been used
in \cite{gutierrez_comparison_2015}. 
We define 
\be\eta=1-p_C,\label{eq:eta}\ee
which is the failure probability of the recovery process of the original logical state.
We will use $\eta$ as the figure of merit we calculate and present in the analysis throughout most of
this paper, except when stated otherwise, as discussed in more detail later.

It is worth highlighting the distinction between this method and
the conventional approach for assessing the success rate of a code (based on a stochastic Pauli model) presented in \seq{sec:intro}, in which every
time a logical operation is applied is categorized as a failure.
The approach that we take here depends on the initial state;
if the noise results in a logical operation that
does not alter the initial state, for example, a logical $Z_L$
applied to the state $|0\rangle_{L}$, it is considered a success.
In this sense, the conventional approach is a worst-case approach deeming every logical operator as a logical error, and this may lead to slight discrepancies between the two methods when applied on the same Pauli noise.

\subsection{Estimating a pseudo-threshold}\label{sec:estimatingpseudo}

A quantum error correction
code is advantageous compared to a physical qubit only for noise parameters such that the probability of an error on the physical
qubit is larger than the probability
of a logical error on the logical qubit with error correction. Assuming a monotonous dependence on one common noise parameter (where the logical qubit is advantageous for small enough values of the noise parameter), the value of this parameter for which the physical and logical error probabilities are equal
determines the pseudo-threshold \cite{gutierrez2016errors, aharonov1997fault}. In our case, this common parameter is the time $t$ between QEC cycles.

We compute the infidelity on a physical qubit by assuming a single physical qubit standing idle and suffering from the 1Q noise terms for time $t$. Since
the 2Q crosstalk does not affect a single isolated physical qubit
when standing idle, defining the pseudo-threshold on a single qubit possibly gives the physical qubit an advantage.
We note that the standard way to compute the
pseudo-threshold must account for the noise affecting  logical states during
the preparation, stabilizer measurements and corrections. However, here
we are within the simplified framework where all of these are noiseless, giving the logical qubit an advantage, so the pseudo-threshold we obtain is merely a code-capacity pseudo-threshold.

In this framework, where the initial
states are simulated explicitly, we choose to examine the pseudo-threshold on the average of
the states in \eq{eq:states to average}.
The averaging is done by computing $\eta$ separately for each state, and then taking the average of all states. Next, we compute the  infidelity for the corresponding physical states, and average over these. Lastly, we find the crossing point of the logical $\eta$ and the physical infidelity.

Nevertheless, for brevity we refer to the ``restricted'' code-capacity pseudo-threshold that we calculate simply as the pseudo-threshold, and note that it is informative for two reasons.
The first reason is that it gives a rough idea of the order of magnitude of the
parameter regime where this error correction is advantageous -- thus roughly
indicating the regime where the validity of the approximations is important for error correction. The
second stems from the fact that the pseudo-threshold itself depends
on the noise model, and so the difference between the pseudo-thresholds
for the different approximations is another way to assess the validity
of an approximation.  

\subsection{The logical density matrix}\label{sec:logical_rho}

 In this subsection we describe how to extract from the noisy state $\rho(t)$ more information than provided by $\eta$ defined in \eq{eq:eta}.
Given a corrupt state $\rho$ and assuming ideal syndrome readout, for each potential syndrome outcome
the state would be collapsed according to that particular syndrome and
the relevant recovery operation applied. To construct the corresponding density matrix the resulting state is multiplied
by the probability associated with that syndrome and the
states for all possible syndromes are summed. This  process yields
the system state  at the conclusion of the
correction procedure,
\begin{equation} \rho \to \rho^P =\sum_s p_s \mathcal{C}_s (1 / p_s) P_s \rho P_s \mathcal{C}_s =\sum_s \mathcal{C}_s P_s \rho P_s \mathcal{C}_s,\label{Eq:CC}\ee
with 
\be p_s = {\rm tr}\{\rho P_s\}, \quad 
P_s = \prod_i \left[\left(\identity + (-1)^{s_i} S_i\right) / 2\right],\ee
where $S_i$ are the stabilizers of the code, and $\mathcal{C}_s$ is the correction operator corresponding to syndrome $s$ (composed of the bits $s_i$, to which corresponds the projector $P_s$). 

For completeness, we note that the stabilizers of the five-qubit code that we investigate are
\begin{align}
S_{1} & =X_{0}Z_{1}Z_{2}X_{3}I_{4},\\
S_{2} & =I_{0}X_{1}Z_{2}Z_{3}X_{4},\\
S_{3} & =X_{0}I_{1}X_{2}Z_{3}Z_{4},\\
S_{4} & =Z_{0}X_{1}I_{2}X_{3}Z_{4},
\end{align}
where $\{X,Y,Z\}$ denote the three Pauli matrices, and the indices indicate the qubit. The Pauli correction operators corresponding to each of the 16 possible syndromes can be found in Tab.~\ref{tab:mod-dec} in \app{sec:mod-dec-details}. The logical Pauli operators are
\begin{equation}
Z_L =Z_{0}Z_{1}Z_{2}Z_{3}Z_{4},\quad
X_L  =X_{0}X_{1}X_{2}X_{3}X_{4}.
\end{equation}

The final step is to evaluate the success rate
of this procedure, achieved by computing the fidelity of the resulting
state with the ideal logical state $\ket{\psi_{0}}$. Moreover, we can calculate also the matrix element of the state between $\ket{\psi_{0}}$ and its orthogonal complement within the logical space, $\ket{\psi_{0}^\perp}$, and define
\begin{equation}
\alpha = 1- \bra{\psi_{0}} \rho\ket{\psi_{0}}, \qquad \beta = \bra{\psi_{0}} \rho\ket{\psi_{0}^\perp}.\label{Eq:alpha_beta}
\ee
To summarize, $\alpha$ is the infidelity of the corrected state with the ideal logical state, or in other words the probability of failure of the correction, and $\beta$ is a measure of coherent logical error that accumulates within the code space. 
We note that $\eta$ which was studied extensively above, is an approximation of $\alpha$ without explicitly performing the corrections in the circuit.
The difference between them stems from eta including only the projection on correctable errors, while alpha is computed on the corrected state, and so could, in general, contain contributions also from non-correctable errors, if following the syndrome projection and correction the resulting state has an overlap with the initial state.
Within the code-capacity model, when the projection onto the syndrome subspaces and the correction are both ideal, $\rho^P$ is guaranteed to be within the code space. 

\section{The approximation Hierarchy}\label{sec:approx_hier}

\begin{figure*}
\centering
\includegraphics[width=0.95\textwidth]{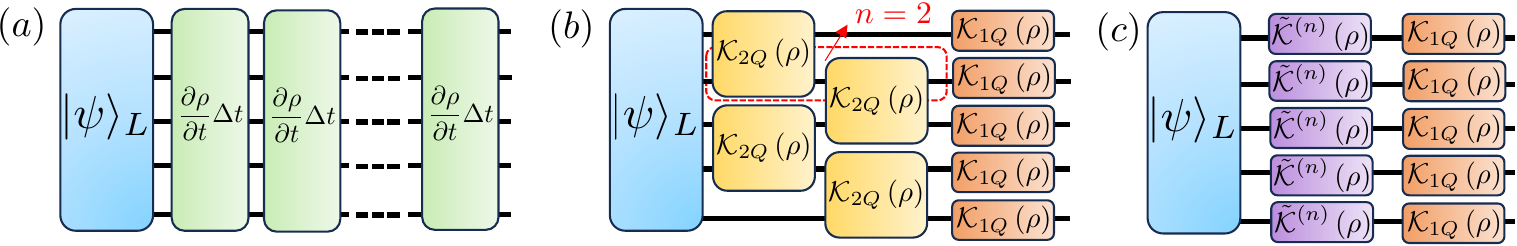}

\caption{(a) A schematic depiction of the dynamical simulation of the underlying noise. The initial logical (pure) state is evolved in small increments of time, governed by the full Lindbaldian and turns into a mixed state.
(b) A schematic depiction of the composite-channel approximation, obtained by solving each term in the Lindbladain separately, grouped into 1Q channels and 2Q channels corresponding to the 1Q terms and 2Q terms in the Lindbladian. We then apply the 2Q channel on every pair of connected qubits in the device, and the 1Q channels on every qubit. For example, in this sketch the qubit marked in red is connected to $n=2$ other qubits.
(c) A sketch of a 1Q approximation, obtained by replacing the 2Q channels by 1Q channels. As discussed in the text, such an approximation is not reliable unless the 2Q noise can be neglected.
\label{fig:circuit sketchs}}

\end{figure*}

In this section we outline the hierarchy of noise channel approximations,
ranging from the most accurate but computationally demanding to the
least precise but easiest to simulate.

\subsection{Dynamical simulation}

The system's state $\rho(t)$ is the solution of the master equation given in \eq{eq:rho}. One numerical method for obtaining $\rho(t)$ proceeds by evolving the solution in small increments of time $\Delta t$, with \be \rho(t+\Delta t) = \rho (t) + \frac{\partial\rho (t)}{\partial t} \Delta t ,\ee as shown in \fig{fig:circuit sketchs}(a).
In practice, we employ an improved numerical method for differential equations \footnote{The Lindblad equation is solved using the Runge-Kutta method, with the numerical relative and absolute tolerances set to $10^{-8}$.}, using the numerical open-source solver \textit{Qiskit Dynamics} \cite{qiskit_dynamics_2023}.

\subsection{The composite-channel approximation}

Each term of the Lindbladian can be easily solved separately, and its operation be represented as a CPTP channel of the Kraus form 
\be \mathcal{E}\left(\rho\right)=\sum_{k}E_{k}\rho E_{k}^{\dagger} ,\qquad\sum_{k}E_{k}^{\dagger}E_{k}=\identity,\label{eq:Kraus}\ee
where for unitary evolution, only one operator is needed. The solution of all terms together is more complex due to the non-vanishing commutation relations between the different terms. Therefore, the first level of approximation we present is the composite-channel approximation. In this approximation every term in the Lindbladian is solved separately to obtain the  corresponding channel, and these channels are applied consecutively. 
This approximation is equivalent to one step of a Trotter expansion, and if we were to increase the number of Trotter steps, it would converge to the dynamical simulation (\apporsm{App:trotter}). 
We now present the channels corresponding to the different terms,  and then go on to compose the full channel out of these.

The term
$H_{i}$ produces the single qubit unitary channel \be \mathcal{U}^z_{\text{1Q}}\left(\rho\right)=U_{1}\rho U_{1}^{\dagger},\label{eq:K1QU}\ee
with
\begin{equation}
U_{1}=e^{-i\frac{1}{2} \left(I-\sigma^{z}\right) ht }=\left(\begin{matrix}1 & 0\\
0 & e^{-iht}
\end{matrix}\right).
\end{equation}
The term $V^{zz}$ gives rise to the two-qubit unitary channel
\be \mathcal{U}^{zz}_{\text{2Q}} \left(\rho\right)=U_{2}\rho U_{2}^{\dagger},\label{eq:K2Q}\ee
with 
\begin{equation}
U_{2}=e^{-\frac{1}{2}i\zeta t \sigma^{z} \sigma^{z} }=\left(\begin{matrix}
e^{-i\zeta t/2} \\
 & e^{i\zeta t/2}\\
 &  & e^{i\zeta t/2}\\
 &  &  & e^{-i\zeta t/2}
\end{matrix}\right),
\end{equation}
that acts on all pairs of qubits that share a connection.

The dissipator part creates a phase and amplitude damping channel \cite{krantz2019quantum}
\be \mathcal{K}_{\text{damping}}\left(\rho\right)=\sum_{k=0}^{2}E_{k}\rho E_{k}^{\dagger},\label{eq:K1QD}\ee
with
\begin{align}
E_{0} & =\left(\begin{matrix}1 & 0\\
0 & \sqrt{1-p_{\rm pd}}\sqrt{1-p_{\rm ad}}
\end{matrix}\right)\\
E_{1} & =\left(\begin{matrix}0 & \sqrt{p_{\rm ad}}\\
0 & 0
\end{matrix}\right)\\
E_{2} & =\left(\begin{matrix}0 & 0\\
0 & \sqrt{p_{\rm pd}}\sqrt{1-p_{\rm ad}}
\end{matrix}\right)
\end{align}
where 
\be 1-p_{\rm ad}=e^{-g_{0}t}, \qquad 1-p_{\rm pd}=e^{-4g_{2}t}.\label{eq:p_ad}\ee
These are related to $T_1$ and $T_2$ of \eqss{Eq:t1t2_def}{Eq:t2_def}.

The full channel is constructed of each of the above-mentioned channels consecutively. The single qubit channels are applied to every single qubit, and the two qubit channels are applied to every pair of qubits that are connected, as depicted in \fig{fig:circuit sketchs}(b). 
When the channels do not commute, we must also specify the order in which these channels are applied, since this may lead to different results.

The conditions for determining whether quantum channels commute can be found in \app{App:commuting_channels}, noticing that these are different from the conditions of commutation of quantum observables. 
Using these conditions we find that $\mathcal{K}_{\text{damping}}$  and $\mathcal{U}^z_{\text{1Q}} $ 
commute and hence can be uniquely composed into a single channel, namely \be \mathcal{K}_{\text{1Q}}\left(\rho\right) =\mathcal{U}^z_{\text{1Q}} \left(\mathcal{K}_{\text{damping}} \left(\rho\right)\right)\label{eq:K1Q}.\ee
However, composing the 2Q channels is more subtle. The channels
$\mathcal{U}^{zz}_{\text{2Q}}$
and $\mathcal{K}_{\text{damping}} $ do not commute, which causes effective qubit dephasing induced by its neighbors' energy relaxation \cite{sdid}. As briefly discussed in \app{sec:XY}, there can be 2Q channels that do not commute when being applied to different pairs that share one common qubit, and this has to be accounted for as well, but this is not the case in \eq{eq:hamiltonian}.

\subsection{The Pauli approximation}\label{subsec:pauli_approx}

Every Kraus operator $E_k$ in \eq{eq:Kraus} can be expanded in the Pauli basis  
$\Pi=\{\sigma | \sigma = \bigotimes_i^N \sigma_i, \sigma_i \in \{ \sigma_i^x, \sigma_i^y, \sigma_i^z, I_i\} \}$
of $2^N\times 2^N$ matrices
as $E_k=\sum_\sigma a_{k\sigma}\sigma$,
and so the channel becomes
\be \mathcal{E} \left(\rho\right) = \sum_{k}\sum_{\sigma\in\Pi} \sum_{\nu\in\Pi} a_{k\sigma}a_{k\nu}^{*} \sigma\rho\nu\label{eq:general channel}.\ee
The Pauli approximation for $\mathcal{E}$ is achieved by discarding all non-diagonal
terms in the sum where $\sigma\neq\nu$, giving
\be\mathcal{P} \left(\rho\right) = \sum_{\sigma} \left(\sum_{k} \left|a_{k\sigma}\right|^{2}\right) \sigma\rho\sigma\label{eq:Pauli},\ee
which is indeed a Pauli channel. 

The main advantage of a Pauli
channel is that it allows one to compute the syndromes without simulating
the state evolution but rather with the parity matrix, as explained in \seq{sec:intro}.
The main physically-driven motivations behind this approximation stems from the influence of the syndrome
measurement on the noise channel, which reduces many non-Pauli terms, explored in more detail in \app{sec:channel_collapse}. 

The most precise way to perform the Pauli approximation is by approximating the whole-system quantum channel. However, this requires constructing the channel which may be numerically as challenging as simulating its action on particular states (or even more). Here we choose a less exact but easier to compute approximation, which consists of approximating each individual channel separately.
The channels $\mathcal{K}_{\text{1Q}}$ and $\mathcal{U}^{zz}_{\text{2Q}}$ are each approximated by 1Q and 2Q Pauli channels, respectively, which are then composed. A more detailed explanation on the different possible ways to perform this Pauli approximation is given in \app{sec:pauli methods}.
We note that the alternative method of approximating the 1Q and 2Q channels together into one 2Q Pauli channel is not a better choice due to the complication arising from the need to ``split'' the 1Q channels into the different 2Q channels in which each qubit is included, which reduces the accuracy of the approximation. 

The Pauli approximation of $\mathcal{U}^{zz}_{\text{2Q}}$ for two qubits indexed (arbitrarily) as 1 and 2 is 
\be
\mathcal{P}_{\text{2Q}}\left(\rho\right)=c_{0}\rho + c_{zz}(\sigma^z_1 \sigma^z_2) \rho (\sigma^z_1 \sigma^z_2),\label{eq:P2Q}
\ee
with
\be
c_{0} = \cos^{2}\left( \zeta t/2\right),\qquad
c_{zz} =\sin^{2}\left( \zeta t/2\right).\label{eq:P2Q_c}
\ee
The Pauli approximation of $\mathcal{K}_{\text{1Q}}$ is
\begin{equation}
\mathcal{P}_{\text{1Q}} \left(\rho\right)= c_{0}\rho+c_{x} \sigma^x\rho \sigma^x+c_{y}\sigma^y\rho \sigma^y+c_{z}\sigma^z\rho \sigma^z,
\end{equation}
with $c_0=1-c_x-c_y-c_z$, and
\begin{align}
c_{x} & =c_{y}  =\frac{1}{4}p_{\rm ad},\\
c_{z} & =\frac{1}{2}-\frac{1}{2}\gamma\cos\left(h_{}t\right)-\frac{1}{4}p_{\rm ad},
\end{align}
where we have defined
\be\gamma =\sqrt{1-p_{\rm pd}}\sqrt{1-p_{\rm ad}}.\label{eq:gamma}\ee
Since the Pauli channels always commute with each other (\app{App:commuting_channels}), for this approximation the order in which we apply the different channels does not matter.

\subsection{1Q approximations} \label{sec:1Q}

Finally, a natural question is whether a 1Q approximation can be devised that is reliable for capturing the 2Q noise, as depicted in Figure \ref{fig:circuit sketchs}(c). Such an approximation -- and in particular if it could be reduced to a 1Q Pauli noise channel -- could be very useful for error correction threshold estimations.
This would require approximating $\mathcal{U}^{zz}_{\text{2Q}}$
by a single-qubit channel, 
since the other channels discussed above are already 1Q channels.
When a two-qubit unitary acts on a separable
state $|\psi_1\psi_2\rangle$, the effect of this on one of the qubits -- disregarding
the second -- is described by a channel with the Kraus operators $E_{k}=\left\langle e_{k}|U|\psi_2\right\rangle $
where $\left\{ |e_{k}\rangle\right\} $ is a basis for the first qubit.
However,
this approach isn't applicable in our case due to the highly
entangled nature of the logical states. 

We tried to create an approximated 1Q noise channel by approximating each qubit to be separable from its neighbors and applying $E_{k}=\left\langle e_{k}|U|\psi_2\right\rangle $, or by identifying the basic actions of the 2Q noise on each qubit, and building a 1Q channel with similar properties.
The 1Q channels we constructed differed qualitatively from the code behavior under 2Q noise at all timescales, and weren't sufficiently adequate even for very short times. 
The details of the 1Q approximations we constructed, and their inferior performance are presented in
 \app{app:1QApproximation}. 
This implies that whenever the 2Q noise magnitude is at least comparable to the other noise terms, it has to be properly accounted for and a 1Q approximation cannot be used. 

\section{Results} \label{sec:approx_results}

In this section we present our main results and highlight the cases where certain approximations hold and what are the mechanisms of their breaking.

\subsection{Dynamical timescales and parameter regimes}

We first note that with 1Q noise only, i.e., for $\zeta=0$, all noise channels commute and so the composite-channel approximation coincides with the dynamical simulation. Examining the Pauli approximation
for different timescales and different initial states, we found it to
be very accurate also, and a few examples are presented in \app{app:1QApproximation}. As mentioned in \seq{sec:lindbladian}, we consider parameters that are identical for all qubits, and taken to be representative of realistic parameters of superconducting-qubit devices, as presented in \app{sec:experiment}. Indeed, the 2Q noise term of idle qubits cannot be neglected with existing superconducting qubits, as can be seen by examining the relevant nondimensional ratios $|\zeta/h|$ and $|\zeta | T_1$. Substituting $h = 2\pi\times 5\,$kHz, $\zeta = 2\pi\times 3\,$kHz and $T_1=150\,\mu$s as typical values obtained for a device of Heron type, we find that the qubits of this device are typically in the regimes of $|\zeta/h |\sim 1$ and $|\zeta| T_1\gtrsim 1$, where the coherent 1Q and 2Q terms are comparable and larger than the incoherent noise. This obviously holds also in typical Eagle devices where $\zeta$ is larger by an order of magnitude. We therefore focus for the rest of this section on the regime of nonnegligible 2Q noise.

Based on these observations, we arrange our results according to three dynamical timescales. We denote these as the short, intermediate, and long times. The long time limit is defined to be that in which $\rho(t)$ approaches a steady state, for $t\gg T_1$. We consider the timescale $t_{\rm L}\sim T_1$ as the border of the long time limit, which for concreteness can be taken to be at $100\mu$s. This limit is explored briefly in subsection \ref{sec:long_time}, and is obviously outside of the regime of useful error correction. 

The intermediate timescale is set by the largest error term that we consider, and can be defined as the time $t_{\rm I}$ obeying $|\zeta| t_{\rm I}\sim 1$. For concreteness, that would be $5\,\mu{\rm s}$ for the typical Eagle value $\zeta=-2\pi\times 30\,$kHz, and it would be $50\,\mu{\rm s}$ with the Heron value of $\zeta=2\pi\times 3\,$kHz.

The short timescale is defined by limit $|\zeta| t_{\rm S}\ll 1$. We can take this timescale to be $1\,\mu{\rm s}$ for Eagle devices and $10\,\mu{\rm s}$ for Heron devices.
The time it takes to apply one 2Q gate on current deployed superconducting devices of IBM Quantum of the Eagle type is of the order of $0.5\,\mu{\rm s}$, and on current Heron devices it is about $0.1\,\mu{\rm s}$, which are both within the short timescale. This timescale is naturally relevant for error correction, when the error rates are relatively low and allow for the duration of at least a few two-qubit gates (or many gates in the Heron case).

The timescales described above and their definitions are summarized in Table \ref{tab:table_times}. In the following subsections we analyze the times at which the various approximations break down quantitatively as well as qualitatively.

\begin{table}
\begin{tabular}{|c|c|c|c|}
\hline 
Timescale & Short & Intermediate & Long \tabularnewline
\hline 
\hline 
Defined by & $|\zeta| t_{\rm S}\ll 1$ &  $|\zeta| t_{\rm I}\sim 1$ & $t_{\rm L}/T_1\sim 1$ \tabularnewline
\hline 
Eagle value & $ 1\,\mu{\rm s}$ & $5\,\mu{\rm s}$ &  $ 100\mu$s \tabularnewline
\hline 
Heron value & $ 10\,\mu{\rm s}$ & $50\,\mu{\rm s}$ &  $100\mu$s \tabularnewline
\hline 
\end{tabular}\caption{A summary of the timescales of error correction dynamics studied in this work. See text for details. \label{tab:table_times}}
\end{table}

\subsection{The long time limit}\label{sec:long_time}

The long-time limit of a noise channel and a single syndrome cycle is not relevant for error correction. Nevertheless, it is informative since it manifests a fundamental difference between the Pauli approximation to the composite-channel and Lindblad dynamics.

\begin{table*}[t!]
\begin{tabular}{|c|c|c|c|c|}
\hline 
$t$ [$\mu{\rm s}$]& Dynamical & Composite, order 1 & Composite, order 2 & Pauli\tabularnewline
\hline 
\hline 
0.5 & $0.015\left(0.005\right)$ & $0.015\left(0.005\right),1.6\times10^{-5}$ & $0.015\left(0.005\right),4.6\times10^{-5}$ & $0.015\left(0.005\right),1.3\times10^{-4}$\tabularnewline
\hline 
1 & $0.055\left(0.016\right)$ & $0.055\left(0.016\right),1.1\times10^{-4}$ & $0.055\left(0.016\right),1.1\times10^{-4}$ & $0.056\left(0.018\right),0.0017$\tabularnewline
\hline 
5 & $0.39\left(0.16\right)$ & $0.39\left(0.16\right),0.005$ & $0.39\left(0.16\right),0.006$ & $0.56\left(0.04\right),0.17$\tabularnewline
\hline 
10 & $0.49\left(0.10\right)$ & $0.49\left(0.09\right),0.008$ & $0.47\left(0.08\right),0.02$ & $0.57\left(0.02\right),0.09$\tabularnewline
\hline 
\end{tabular}\caption{Table of values of $\eta$ for different approximations and different noise durations. The noise parameters are given in \eqss{eq:EagleParams1}{eq:EagleParams2}. 
The values presented in this table are $\left\langle \eta\right\rangle \left(\text{std}\left(\eta\right)\right),\sqrt{\left\langle \left(\eta-\eta_{\rm{D}}\right)^{2}\right\rangle }$
where  the average $\left\langle .\right\rangle$ and standard deviation $\text{std}\left(.\right) $ are taken over the six Pauli eigenstates as initial
states [\eq{eq:states to average}] and $\eta_{\rm{D}}$ is $\eta$ under the dynamical simulation of the noise on each initial state. The term $\sqrt{\left\langle \left(\eta-\eta_{\rm{D}}\right)^{2}\right\rangle }$ is the square root of the average of the square of the difference between $\eta$ of the approximated noise and $\eta$ of the dynamical simulation of the noise. It gives us information on the average of the discrepancy in $\eta$ due to the approximation of the noise. 
The composite-channel approximation does very well, while the Pauli approximation is good for short times but differs significantly for longer time scales, for which, in this example, it overshoots the failure rate. 
\label{tab:table correctability}}\end{table*}

For the Kraus channels in the long-time limit, we have $\rho(t\to\infty)\to |0\rangle\langle 0|^{\otimes N}$ for any finite value of $T_1<\infty$. However, for the Pauli channels in the long-time limit, we have $\rho(t\to\infty)\to \identity/2^N$, since the 1Q Pauli approximation reduces to the fully depolarizing channel in this limit. 
This difference in the steady state necessarily implies that at least for $t\gtrsim T_1$ (if not earlier), we can {\it a-priori} expect any Pauli approximation to break
\footnote{The steady state being the fully mixed state results in $\eta \to 1/2$ in the Pauli approximations for any initial state. In contrast, noticing that the state $|0\rangle_L$ ($|1\rangle_L$) is a superposition over the 5-bit strings with even (odd) number of 1's, we can calculate the fidelity and correctability of the state in the composite-channel approximation at $t\to\infty$, which gives us $\eta\to 3/8$ for $|0\rangle_L$, $\eta\to 5/8$ for $|1\rangle_L$ and $\eta\to 1/2$ for $|+\rangle_L$.}.

\subsection{Dominant ZZ crosstalk noise}

We begin by taking noise parameters compatible with the IBM Eagle device (\app{sec:experiment});
\be h=-2\pi\times 5\,{\rm kHz},\qquad  \zeta=-2\pi\times 30\,{\rm kHz},\label{eq:EagleParams1}\ee
\be\quad T_{1}=150 \,\mu{\rm s}, \qquad T_{2}=100 \,\mu{\rm s}.\label{eq:EagleParams2}\ee
In \fig{fig:cusco}(a) we show the average of $\eta$ over the initial states
in \eq{eq:states to average}, in the short-time regime.
Since the separate channels in the composite approximation do not commute, the order in which they are applied matters. In \fig{fig:cusco}, the label ``composite 1'' refers to applying the 1Q channels before the 2Q channels, and ``composite 2'' implies the reverse order.
The results are very similar for every initial state separately. It can be seen that all approximations do very
well. 
In \fig{fig:cusco}(b), we examine times up to the intermediate timescale, focusing on the initial state $|0\rangle_{L}$, for which the Pauli approximation differs significantly and qualitatively, but for long-enough times, also the composite-channel approximation
deviates.

\begin{figure}
\centering
\includegraphics[width=0.48\textwidth]{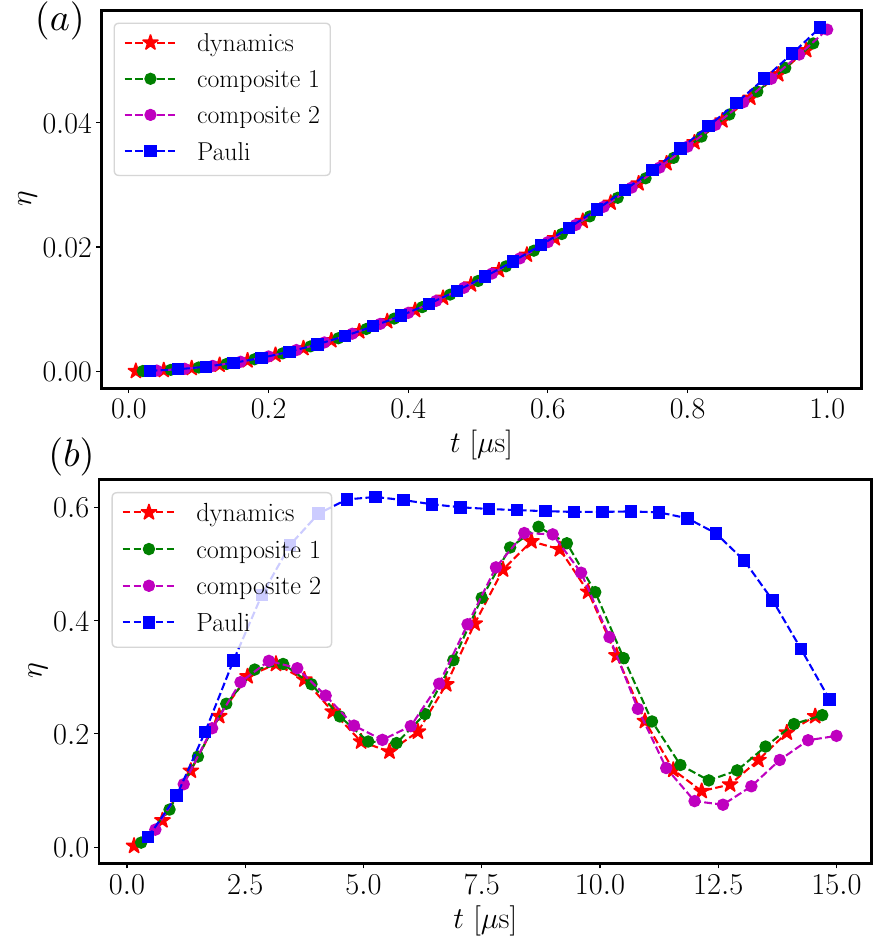}
\caption{\label{fig:cusco}The failure probability after error recovery -- $\eta$ of \eq{eq:eta} -- vs.~$t$, the duration of the noise channel, and the noise parameters in \eqss{eq:EagleParams1}{eq:EagleParams2}, with dominant crosstalk noise. (a) An average over the initial states in \eq{eq:states to average}, in the short-time regime, showing that all approximations are identical. (b) The initial state is $|0\rangle_L$, shown up to the intermediate timescale, when the Pauli approximation breaks, and also the composite-channel approximation loses accuracy (shown for two possible orderings, see the text for details).
}
\end{figure}

Table \ref{tab:table correctability} compares the accuracy of the different approximations, by showing the mean and standard deviation when averaged over the initial states in \eq{eq:states to average}.
In order to express the dependence of the success rate on the initial state, we also present the standard deviation over initial logical states, of the difference between $\eta$ of the dynamical simulation and that of the approximation. We note that the pseudo-threshold in this case is very low (\app{App:pseudo}) since the noise is dominantly 2Q noise and when being compared with a physical (single) qubit, the single qubit does not suffer from this noise.

This data strengthens the observations made above, that for the studied parameters the composite-channel approximation is a very good approximation up to intermediate times, while the Pauli approximation works well only for relatively short time-scales.
Regarding the composite-channel approximation losing accuracy, this results from the noncommutativity of the 2Q crosstalk channel with amplitude damping. The relevant parameter is $|\zeta| t \times t/T_1$ that remains a relatively small parameter (reaching $\approx 0.28$ for $t=15\,\mu$s at the right boundary of \fig{fig:cusco}). This is a general observation about the nature of the noise model, and the effect of noncommuting terms in the approximation is further explored in \app{sec:composite big t1}.

In the current example, the 2Q noise is the largest term, and is one that cannot be corrected with the employed standard (1Q-noise) decoder of the 5Q code. This results in a near-parabolic dependence on $t$ in \fig{fig:cusco}(a) consistent with \eqss{eq:P2Q}{eq:P2Q_c}. The Pauli approximation overestimates the failure rate when leaving the short time regime, plausibly since it does not describe the coherent dynamics accurately enough (when in this timescale, coherently accumulated phases can lead to cancellations of errors that are impossible in the Pauli description). In order to gain more insight into this, we consider in the next subsection a decoder tailored to the noise.

\subsection{Modified decoders}\label{sec:mode_dec_details}

The five-qubit code with the standard decoder is a distance-three code, i.e., it can successfully correct any single 1Q error.
However, since the noise model includes also 2Q ZZ terms, when the probability of those 2Q errors is larger than that of (at least some of) the 1Q errors, it could be beneficial to sacrifice the correction of 1Q errors in order to gain the ability to correct the more prevalent 2Q errors. In this case the decoder can be best constructed to account also for qubit connectivity.
In \app{sec:mod-dec-details} we present in detail how the standard five-qubit code decoder can be modified in order to correct one $Z_iZ_j$ error on any pairs of qubits. This is achieved without changing the code states or syndromes but rather reinterpreting them, by sacrificing the ability to correct single qubit $X$ or $Y$ errors (while keeping the correction of all $Z$ errors untouched).
We denote this modified decoder as the full-connectivity (FC) ZZ decoder.

\begin{figure}
\centering
    \includegraphics[width=0.48\textwidth]{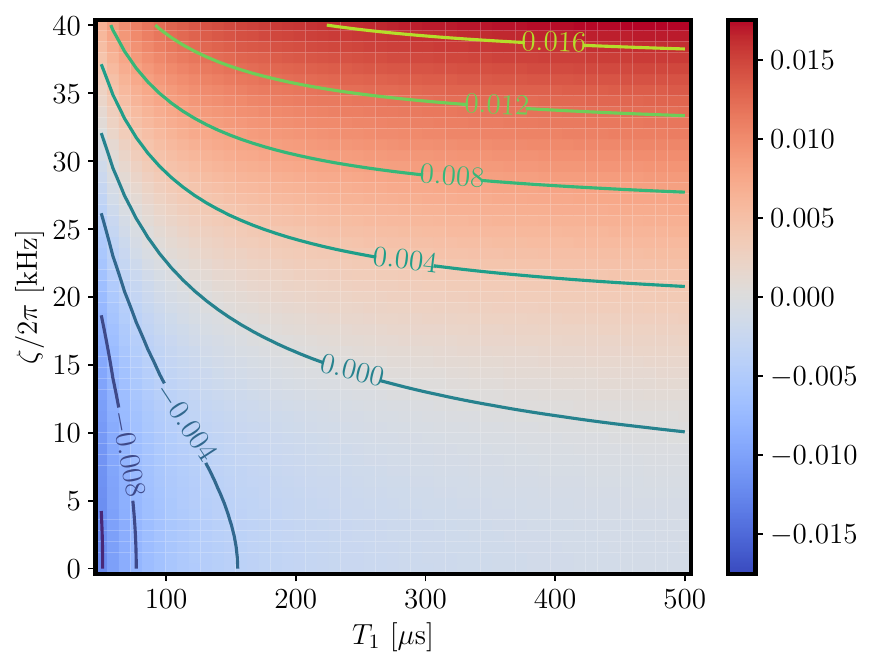}
\caption{\label{fig:modified decoder}The difference between the logical error when the FC ZZ decoder is used vs.~the standard decoder, with  ZZ crosstalk noise acting with all-to-all connectivity. Here $h=0$, $T_2=2T_1$, and the initial state $|+\rangle_L$ is evolved for $t=0.5\,\mu$s using the dynamical simulation.
Positive values (colored red) imply that the modified decoder is beneficial, and this can be seen to hold for a wide range of parameters in the regime of dominant crosstalk. }
\end{figure}

In \fig{fig:modified decoder} we present the difference between $\eta$ with the FC ZZ decoder and with the standard decoder, showing that the FC ZZ decoder is beneficial in a wide range of experimentally relevant parameters. 
Specifically, the parameter values of \eqss{eq:EagleParams1}{eq:EagleParams2} fall in the regime where the modified decoder is advantageous. Therefore, applying the FC ZZ decoder to the dynamics with the same parameters as in \fig{fig:cusco}(a), the FC ZZ decoder performs much better than the regular decoder, obtaining much lower values of $\eta$, as can be seen in \fig{fig:eagle_mod}. Since Z and ZZ errors are corrected at the leading order, the dominant errors are single-qubit flips ($T_1$), with a linear time dependence as in \eq{eq:p_ad}. The Pauli approximation begins to diverge from the dynamical simulation at $t\approx 0.7 \mu$s, underestimating the failure rate. In other words, even at lower failure rates and when the code is adapted to the dominant noise terms, the Pauli approximation could lose its accuracy already in the short time regime, and not provide even an upper bound on the logical error.

In \app{sec:mod-dec-details} we present in more detail 
results for the case of qubits in a ring topology, using an adequate choice of a slightly different decoder. We also show that with a decoder that does not match the qubits' connectivity, the failure rate is increased.

\begin{figure}
\centering
    \includegraphics[width=0.48\textwidth]{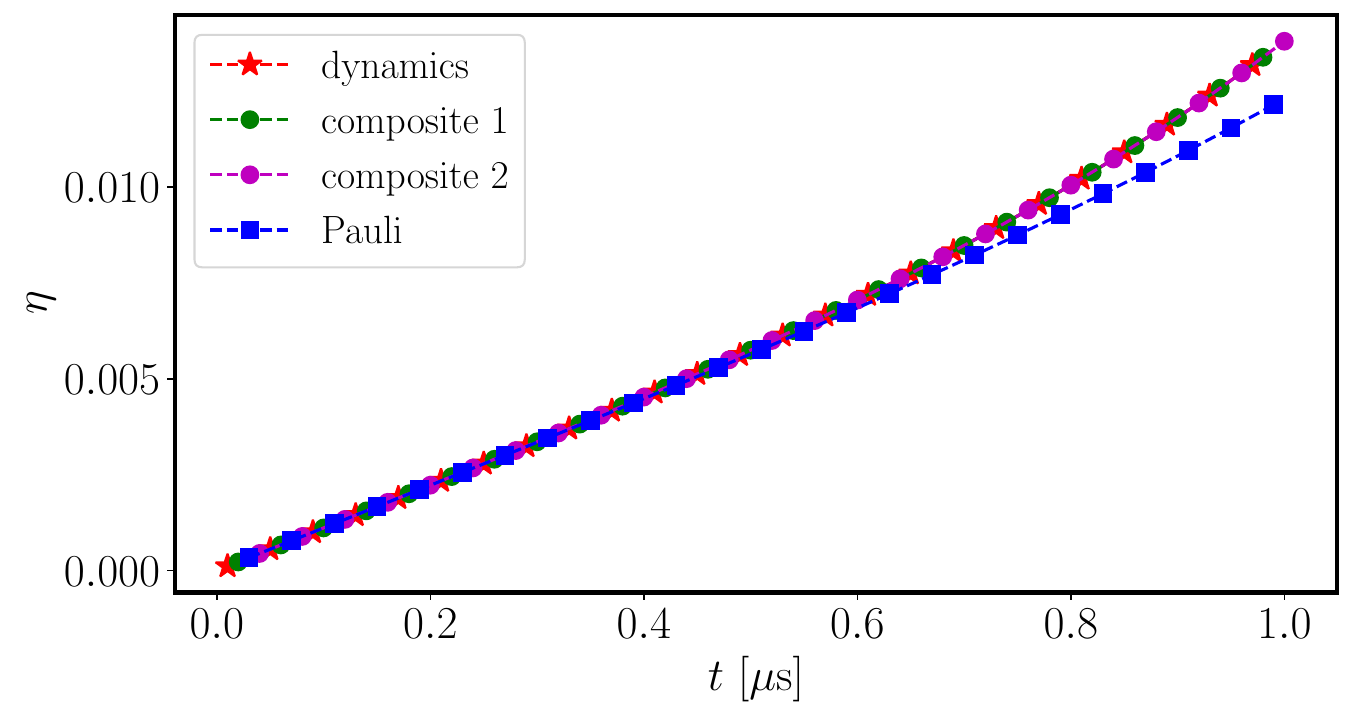}
\caption{\label{fig:eagle_mod}The failure probability after error recovery -- $\eta$ of \eq{eq:eta} -- vs.~$t$, for $t$ the time of the noise channel,  and the noise parameters in \eqss{eq:EagleParams1}{eq:EagleParams2}, with dominant crosstalk noise, and the modified FC ZZ decoder. An average over the initial states in \eq{eq:states to average}, in the short-time regime, shows that the FC ZZ decoder performs much better than the regular decoder, even though the Pauli approximation begins to diverge from the dynamical simulation already at $t\approx 0.7 \mu$s.
}
\end{figure}

\subsection{Purely ZZ crosstalk}\label{sec:PureZZ}

\begin{figure}
\centering
    \includegraphics[width=0.48\textwidth]{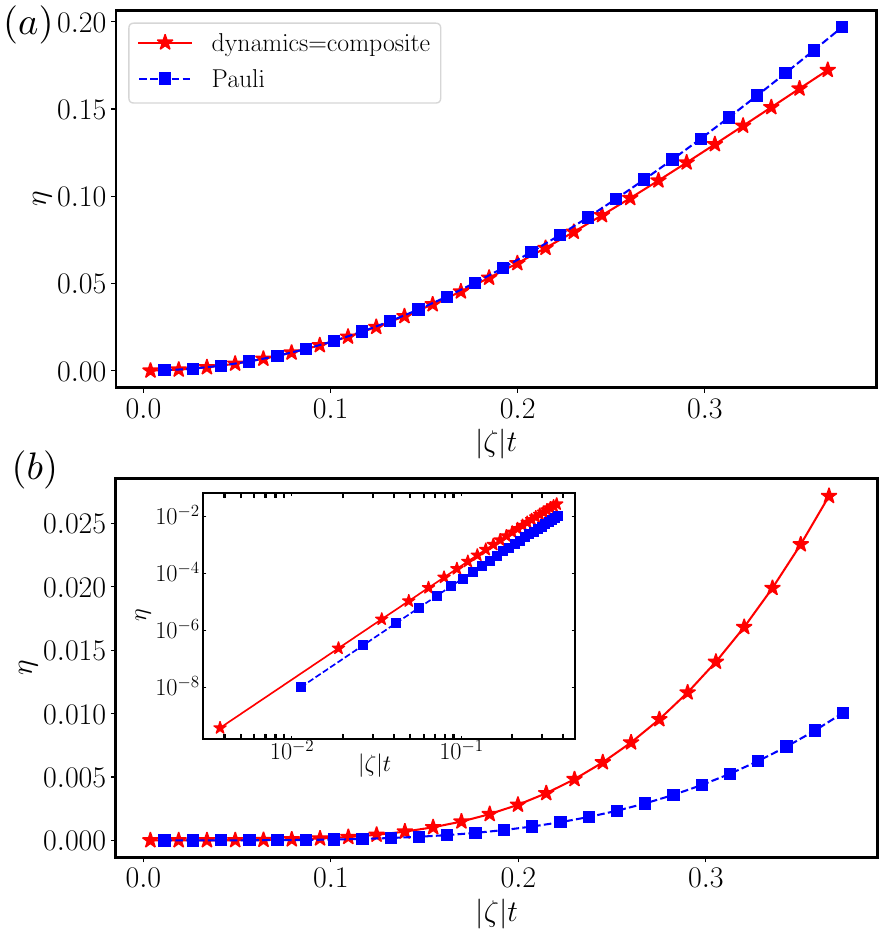}
\caption{The failure probability $\eta$ vs.~$|\zeta|t$, for $t$ the time of the noise channel, with only ZZ crosstalk noise, of magnitude $|\zeta|$ [\eq{eq:2QParams}]. The composite-channel approximation is exact and identical to the full dynamical solution in this case. The curves are an average over the six logical Pauli eigenstates [\eq{eq:states to average}] in the short-time limit, with (a) the regular decoder, and (b) the FC ZZ decoder (the inset showing the same data using log-log axes). We can see that the latter decoder outperforms the former (by about an order of magnitude), as expected. The Pauli approximation holds until $|\zeta| t\approx 0.2$ for the regular decoder, while it doesn't hold for any value of $|\zeta| t$ for the FC ZZ decoder. See the text for a detailed discussion.\label{fig:2Q only}}
\end{figure}

In order to gain a deeper understanding of the Pauli approximation failure we turn to analyze the case where the sole source of noise is the 2Q term $V^{zz}$, i.e., we set 
\be h=0,\qquad T_{1}=T_{\phi}=  T_{2}=\infty.\label{eq:2QParams}\ee
In this case,
since the noise terms on all pairs of qubits commute, the composite-channel approximation is equivalent to the dynamical simulation, and, as stated above, our focus is on the validity of the Pauli approximation. The errors only depend on the combination
$\zeta t$, and in the current case with $h=0$, we observe that the dynamics are independent of the sign of $\zeta$.

For the regular decoder, we see in \fig{fig:2Q only}(a) that the Pauli approximation does very well up to $|\zeta |t \approx 0.2$ on the average of the initial states. For concreteness, the value $|\zeta| t=0.2$ is obtained at $t\approx 1\,\mu$s with $\zeta=-2\pi\times 30\,$kHz, while for $\zeta=2\pi\times 3\,$kHz it is reached at $t\approx 10\,\mu$s.
We note that there are large differences between initial states, with the approximation holding for much larger values for some initial states.

The FC ZZ decoder, presented in \fig{fig:2Q only}(b), outperforms the regular decoder for small values of $|\zeta| t$, as expected and is evident by the failure rates smaller by an order of magnitude.
However, it can be seen that for this decoder, the Pauli approximation does not hold for any value of $|\zeta| t$. An expansion analysis in fact shows that in this case, the Pauli approximation constructed from the 2Q noise parameters (and corresponding generators) underestimates the error by a factor of 3 (up to perturbative corrections) for some states and any noise amplitude $\zeta$. This factor is derived in \app{sec:PauliCounting} where it is seen to stem from unitary evolution terms in second order that are neglected in the construction of the Pauli approximation.
As discussed in \app{sec:PauliCounting} and also further in \seq{Sec:LogicalDM}, this scaling factor of the crosstalk terms in the Pauli channel in the general case depends on the details of other noise terms 
(and the particular state, and decoder), and therefore needs to be derived numerically by comparison of the failure rate to the dynamics.

\subsection{Lower crosstalk noise}

In this subsection we explore the regime of crosstalk smaller by an order of magnitude together with a comparable detuning error, which is compatible with a Heron device,  (\app{sec:experiment});
\be h=2\pi\times 5\,{\rm kHz},\qquad \zeta=2\pi\times 3\,{\rm kHz},\label{eq:HeronParams1}\ee 
and for the dissipative parameters we keep as before,
\be 
T_{1}=150\,\mu {\rm s}, \qquad T_{2}=100\,\mu {\rm s}.\label{eq:HeronParams2}\ee 

For these parameters, as can be seen in \fig{fig:modified decoder}, the regular decoder is advantageous, and so we will only use the regular decoder throughout this section.
In \fig{fig:heron} we present $\eta$ for the different approximations,
on the average of the initial states in \eq{eq:states to average}, and note that the individual states behave similarly. We can see that the
composite-channel approximation, as well as the Pauli approximation, are very good
approximations within the short-time limit of the dynamics, though they begin to lose accuracy at $t\sim 6 \mu$s, consistent with expected (small but not negligible) contributions from coherent accumulation of phases (1Q and 2Q) and noncommuting terms.

We have also calculated the pseudo-thresholds (within the code capacity noise framework, as explained in \seq{sec:estimatingpseudo}) for these noise parameters, for the dynamical simulation and the approximations. The 
pseudo-thresholds for the averaged state are $6.21$ $\mu$s for the dynamical simulation,
$6.71$ ($5.65$) $\mu$s for the composite-channel order 1 (order 2) approximation, and $5.73$ $\mu$s
for the Pauli noise. More details can be found in \app{App:pseudo}.
They are especially relevant since a 2Q gate duration is of order $0.1$ $\mu$s on those devices, potentially allowing for enough rounds of 2Q
gates on the qubits before reaching the pseudo-threshold. In addition,
we see that for these parameters, the pseudo-threshold varies to some extent between the different approximations. 

\begin{figure}
\centering
    \includegraphics[width=0.48\textwidth]{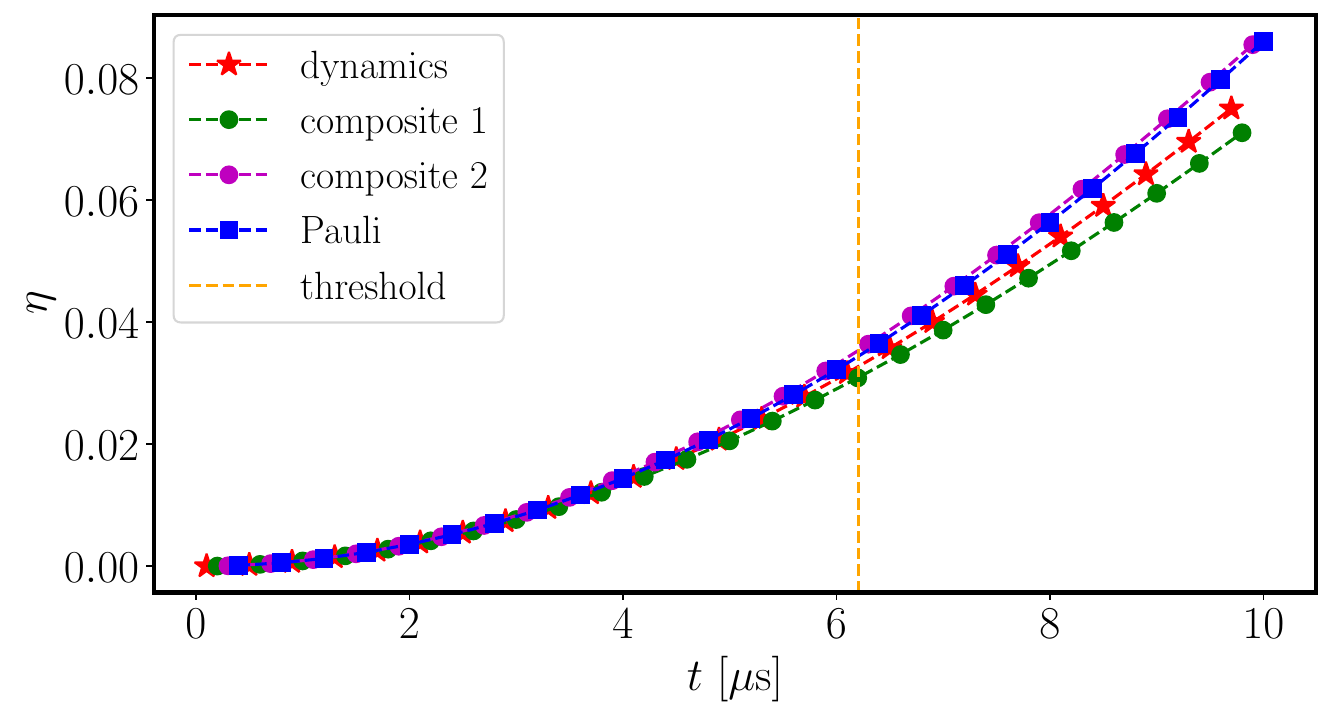}
\caption{\label{fig:heron} The failure probability $\eta$ vs.~$t$ shown for the average over initial states in \eq{eq:states to average}, with the noise parameters in \eqss{eq:HeronParams1}{eq:HeronParams2}, corresponding to lower crosstalk (as compared with \fig{fig:cusco}), together with a comparable detuning error. The time axis covers the short-time regime and the approximations are seen to begin diverging towards its limit.
}
\end{figure}

\subsection{Nonsymmetric reference frame}

\begin{figure}
\centering
    \includegraphics[width=0.48\textwidth]{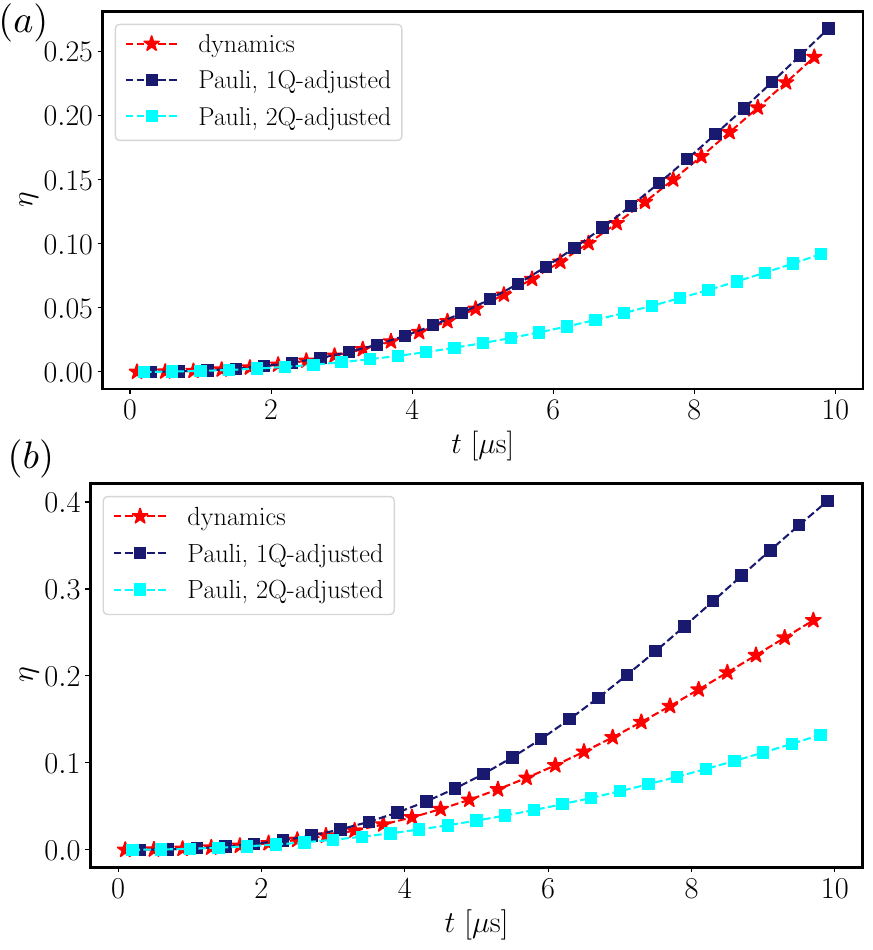}
\caption{\label{fig:non-sym} The failure probability $\eta$ vs.~$t$, where the noise parameters are $\tilde{h}=2\pi\times 5$ kHz, $\zeta=2\pi\times 3$ kHz, $T_1=150$ $\mu$s and $T_2=100$ $\mu$s, in the short time scale. The initial states  are: (a) $|+\rangle_L$, (b) $|0\rangle_L$. Performing the Pauli approximation via adjusting the 1Q term could be more accurate in some cases but is not decisively so, and it might both overestimate and underestimate the failure rate, already in the short time regime.
}
\end{figure}

The Hamiltonian model that we considered so far [\eqss{eq:hamiltonian}{Eq:1Q_ham}] corresponds to a frame of reference such that each qubit's frequency is set at the mean of the positive and negative state-dependent crosstalk shifts, which we will call the symmetric frame of reference.
However, in practice the frequencies are often calibrated when all neighbors of each qubit are in the ground state \footnote{This is the case with current IBM Quantum devices deployed on the cloud \cite{IBMQuantum}.}. This calibration sets a different frame, with an effective nonsymmetric 2Q crosstalk of the form
\begin{equation}\label{Eq:non_sym}
\tilde{V}_{ij}^{zz}=\frac{1}{2}{\zeta}_{ij} (I - \sigma^z_{i}) ( I - \sigma^z_{j}),
\end{equation}
with a corresponding Hamiltonian,
\begin{equation}
\tilde{\mathcal{H}}/\hbar =\sum_{i} \frac{1}{2}\tilde{h}_{i} \left(I_i-\sigma^z_{i}\right) +\sum_{\left\langle i,j\right\rangle }\tilde{V}^{zz}_{ij},\label{eq:hamiltonian2}
\end{equation} 
and the process of calibrating the qubits effectively consists of minimizing the detuning error $\tilde h_i$ in this frame.

A direct use of $\tilde{\mathcal{H}}$ in the approximations would implement this reference frame by using the amended 2Q noise term of \eq{Eq:non_sym}, which we denote as ``2Q-adjusted''. It is also possible to represent the nonsymmetric frame of $\tilde{\mathcal{H}}$, by noticing that it is equivalent to $\mathcal{H}$ in \eq{eq:hamiltonian} if we choose for each qubit ${h}_i = \tilde{h}_i - \sum_{\langle j\rangle}\zeta_{ij}$, where the sum is over the qubit's neighbors. This approach directly modifies the 1Q noise term and we denote it as ``1Q adjusted''.
These two representations are identical when we are interested only in the dynamical simulation or the composite-channel approximation.

However, when we are interested in the Pauli approximation, these two representations for the nonsymmetric reference frame are different. This stems from the fact that the Pauli approximation is done on the 1Q noise terms and 2Q noise terms separately. 
These differences can be seen in \fig{fig:non-sym}, for 
$\tilde{h}=2\pi\times 5\,$kHz, ${\zeta}=2\pi\times 3\,$kHz, $T_{1}=150$ $\mu$s and $T_{2}=100$ $\mu$s.
In \fig{fig:non-sym}(a) we see that for the initial state $|+\rangle_{L}$ in the ``2Q-adjusted'' representation, the Pauli approximation begins to deviate already at
$t=0.3$ $\mu$s, while in the ``1Q-adjusted'' representation, the Pauli
approximation remains accurate up to longer times.
The initial state $|0\rangle_L$, on the other hand is
significantly different, in that the Pauli approximation does not perform well regardless of the representation, as can be seen in \fig{fig:non-sym}(b).
In total, it seems that for these noise parameters, the 1Q-adjusted representation is somewhat beneficial for the Pauli approximation, although again, we see that the Pauli approximation is sensitive to the specific noise and state parameters.

\subsection{Dynamics of syndrome cycles}\label{Sec:LogicalDM}

In this subsection we go beyond the picture studied so far where the failure rate of codes has been presented for a noise channel acting for varying durations on the system, following which an ideal procedure of syndrome extraction and recovery has been assumed. We also depart in this section from the quantity $\eta$ that has been our measure of the recovery failure rate so far, and we consider matrix elements of the logical density matrix following repeated error correction syndrome cycles.

We employ the modified FC ZZ decoder and we restrict the noise terms to those compatible with this decoder,
\be h=2\pi\times 5\,{\rm kHz},\qquad \zeta=2\pi\times 8\,{\rm kHz},\label{eq:FCZZParams1}\ee 
and for the dissipative parameters,
\be 
T_{1}=\infty\, \qquad T_{2}=200\,\mu {\rm s}.\label{eq:FCZZParams2}\ee 
Therefore all noise terms are correctable (to first order) by the code. We assume a syndrome cycle of a duration of 1$\mu s$ (which is roughly the minimum possible with state-of-the-art superconducting qubit devices), during which the noise acts and after which an ideal syndrome projection and Pauli correction are applied (instantaneously), as discussed in \seq{sec:logical_rho}. The resulting density matrix following the cycle [$\rho^P(t)$ of \eq{Eq:CC}] is used as the initial state for the next cycle. 

\begin{figure}
\centering
    \includegraphics[width=0.49\textwidth]{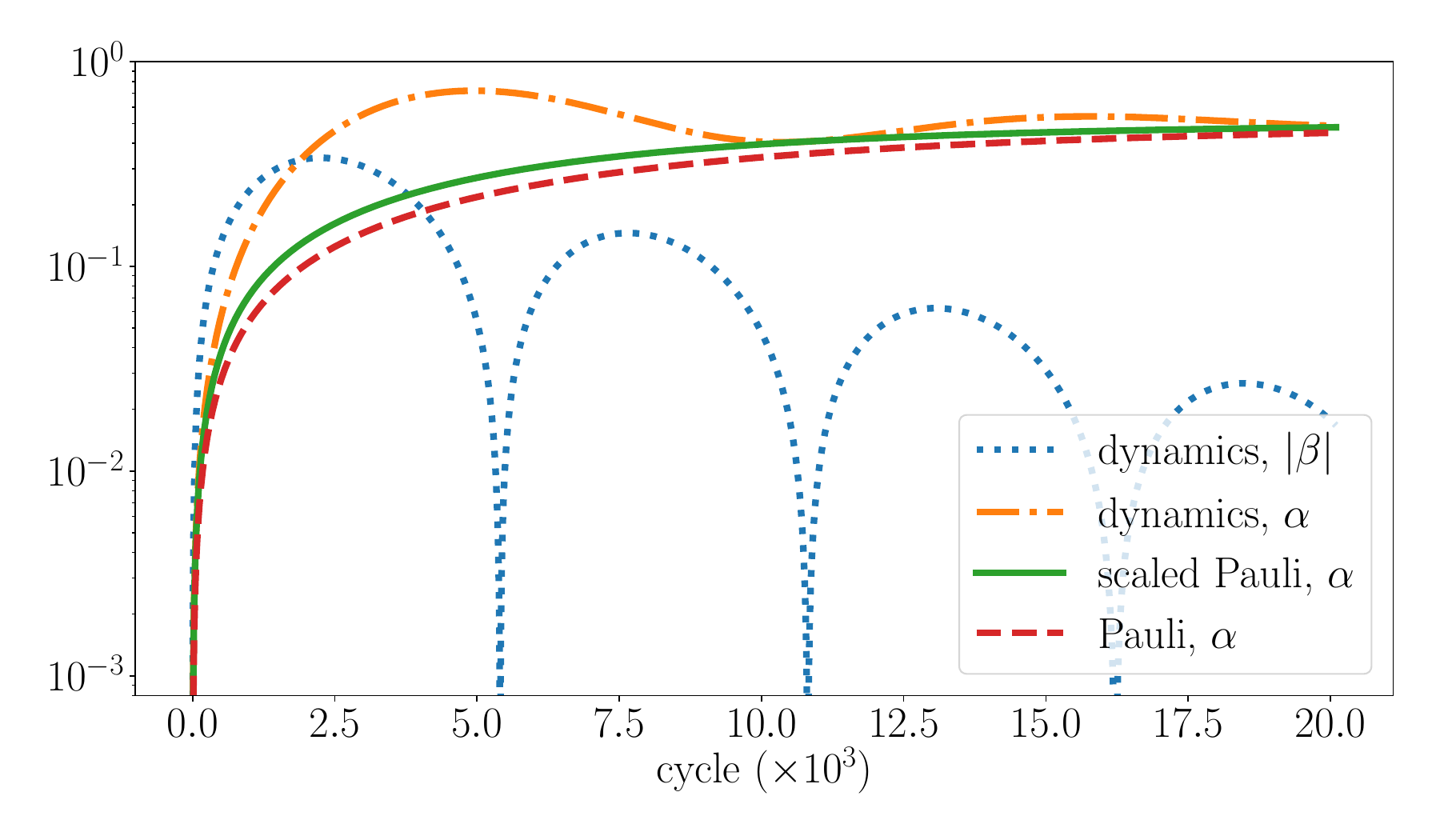}
\caption{\label{fig:cycles} Matrix elements of the logical density matrix, within the dynamical simulation and the Pauli approximations, for continuously repeated syndrome cycles of noise, syndrome projection and recovery, of duration 1$\mu s$ per cycle. The input logical state is $\ket{+}_L$ and the plotted parameters $\alpha$ -- infidelity and $\beta$ -- coherent matrix element, are defined in \eq{Eq:alpha_beta}. Dynamical noise parameters are given in \eqss{eq:FCZZParams1}{eq:FCZZParams2}. The scaled Pauli approximation has been scaled to make the failure rate after one syndrome cycle identical to the dynamics, but it can be seen to miss the evolution of the system. See the text for the details of the model.
}
\end{figure}

Figure \ref{fig:cycles} presents the infidelity of the resulting state with the ideal initial state $\ket{+}_L$ and the coherent matrix element with $\ket{-}_L$ as defined by $\alpha$ and $\beta$ in \eq{Eq:alpha_beta}, for an evolution over 20000 cycles obtained from the dynamical simulation. 
The value of $\alpha$ is shown for a Pauli approximation of the noise channel (based on its duration of 1$\mu s$ and subject to identical projection and correction cycles), and also for a scaled Pauli channel, obtained using the procedure discussed in \seq{sec:PureZZ}; We scale the 2Q Pauli crosstalk strength by a heuristic factor ($\approx 1.149$) such that $\alpha$ of the dynamics and the scaled Pauli approximation agree at the end of the first syndrome cycle (within a fractional precision of $\lesssim 0.6\times 10^{-3}$). 

Nevertheless, this heuristic factor is not sufficient to fix the Pauli approximation and as the figure shows, after about 200 cycles the state in the dynamical simulation begins to deviate significantly from the scaled Pauli approximation, which significantly underestimates the error (and remains constantly below it).
Notably, the value of $\alpha$ appears to oscillate towards a possible steady state (which is seen to be 1/2), and the coherence decays.
It is plausible that the coherent phase accumulating within the logical subspace (that is not modeled at all within the Pauli approximation) contributes to this different dynamical evolution. The existence of such rich dynamics on the scale of a large number of syndrome cycles cannot be captured within the stochastic Pauli model, which carries no coherence between cycles.

\section{Summary and Outlook}\label{sec:closing}

This work is focused on the importance of  physically-motivated noise
dynamics when determining the success rate of an error correction
code to recover logical states subject to noisy dynamics. 
The Lindblad model studied in this work includes the dominant idle errors in superconducting quantum computing systems \cite{shirizly2024dissipative}, or can be modified in order to better model those \footnote{We note that the effect of frequency shifts caused by stochastic charge-parity fluctuation in superconducting qubits \cite{shirizly2024dissipative}, can be absorbed into each qubit's $h_i$ value and averaged over the possible frequencies.}. The Markovian decoherence and qubit crosstalk that we focused on here constitute in many physical systems the leading-order noise terms \cite{cao2023generation, baumer2023efficient, graham2022multi,yang2022realizing,monz201114, monroe2021programmable}. As noise is reduced on qubits and control techniques improve, non-Markovian modeling of the environment may become important \cite{white2020demonstration, PhysRevApplied.17.054018,white2023unifying, PhysRevApplied.21.024018, velazquez2024dynamical, varona2024lindblad} and it would be interesting to see how approximations of error-correction dynamics perform.

As mentioned in \seq{sec:lindbladian}, in some quantum devices the qubit parameters vary from qubit to qubit. The natural question that arises is how an inhomogeneity in the dynamical parameters would affect the conclusions of our study. In \app{sec:inhomogeneous} we have considered the five-qubit code dynamics with some spread in the parameters (relevant to superconducting qubits), and we see that the error-correction dynamics of the spatially inhomogeneous system could still be qualitatively similar to the dynamics observed with uniform qubits.

Overall, we have determined that for a Lindbladian of the form presented in \seq{sec:lindbladian}, the composite-channel approximation (constructed explicitly to account for 2Q noise) is a very good approximation for relevant noise parameters and the time scales that we study. We show that the composite-channel approximation holds up to time scales for which the non-commutativity of the individual channels becomes nonnegligible. This approximation will also hold for other systems with similar ratios between noise parameters, by scaling the duration of the noise channel accordingly. 
Modeling other coupling terms can be more subtle due to noncommuting terms possibly contributing in shorter time scales, as in the case of an XY coupling term, which we briefly consider in \app{sec:XY}. This understanding would be an important guideline when modelling more general setups than the one considered here, e.g., when incorporating the dynamics during gates involving parity-check qubits. We have also considered the effect of choosing a reference frame, such as the non-symmetric reference frame often used with superconducting qubits hardware, the different ways the Pauli approximation can be performed in this case, and the affect it has on the performance of the approximation.

When analyzing the properties of a code, simplified 1Q Pauli noise channels are often assumed in order to facilitate the standard analysis. Although for 1Q noise the Pauli approximation holds, we did not manage to find a 1Q approximation that captures the success rate of the code under 2Q ZZ crosstalk, even for very short times, as explained in \app{app:1QApproximation}. Studying the 2Q Pauli approximations we find that its applicability is limited to the low-noise regime (short time scales), which is the most important regime for error correction. However, it is not a controlled approximation since it can be sensitive to the details of the combination of the state and decoding scheme (in addition to interactions between noise terms), and we pointed out significant mechanisms of its failure. We saw several scenarios where the Pauli noise underestimates the logical error rate and hence cannot be taken as a pessimistic model for the error correction rates.

Beyond the feasibility of simulating large code families, the Pauli approximation is prevalent in error correction studies because an essential property of quantum error correction is that the syndrome measurement changes the noise channel, and many of the non-Pauli terms disappear this way, as discussed in \app{sec:channel_collapse}. 
However, since some non-Pauli terms still survive even after the syndrome measurement, this approximation is not exact.
In fact, we have seen that after a large number of syndrome cycles, the success rate predicted by the Pauli approximation as we have constructed it misses the underlying dynamics and there does not seem to be an immediate way to remedy this since the culprit seems to be the accumulation of small phase contributions that are inaccessible with Pauli channels.
Accounting for imperfect syndrome cycles using a circuit model could be expected to introduce new complexities -- e.g., the accumulation of a coherent phase between states in the code space and outside of it, which was not possible in the current model.

Therefore, a fundamental open question is how our observations would change with circuit-based non-Pauli dynamics of error correction, and what would be the relevant effects with larger codes with an increasing distance, and some randomness in the parameters. It will be interesting to identify which properties are common to all codes, and which properties are unique to different codes. In particular, how the error correction threshold and logical error rates scale and change when treated within
the different approximations, and whether the possible accumulation of coherent logical errors across syndrome cycles observed in \seq{Sec:LogicalDM} remains relevant with codes of higher distance. This direction could be nontrivial
to pursue since it would require the simulation of the error correction
procedure with large codes. This
may be difficult with a dynamical simulation, but perhaps feasible with the composite-channel approximation, and in particular, using tensor-network methods that are suitable for the relevant states \cite{shirizly2024dissipative}. 
An understanding of this question may be useful to shed a light on the feasibility of error correction under realistic noise and could possibly contribute to its realization.

\section*{Acknowledgements}
We thank Luke Govia, Maika Takita, James Wootton, Eliahu Cohen and Dorit Aharonov for very helpful feedback.
Research by H.L. and L.S. was sponsored by the Army Research Office and was accomplished under Grant Number W911NF-21-1-0002. The views and conclusions contained in this document are those of the authors and should not be interpreted as representing the official policies, either expressed or implied, of the Army Research Office or the U.S. Government. The U.S. Government is authorized to reproduce and distribute reprints for Government purposes notwithstanding any copyright notation herein.

\appendix

\begin{figure*}
    \centering
    \includegraphics[width=0.99\textwidth]{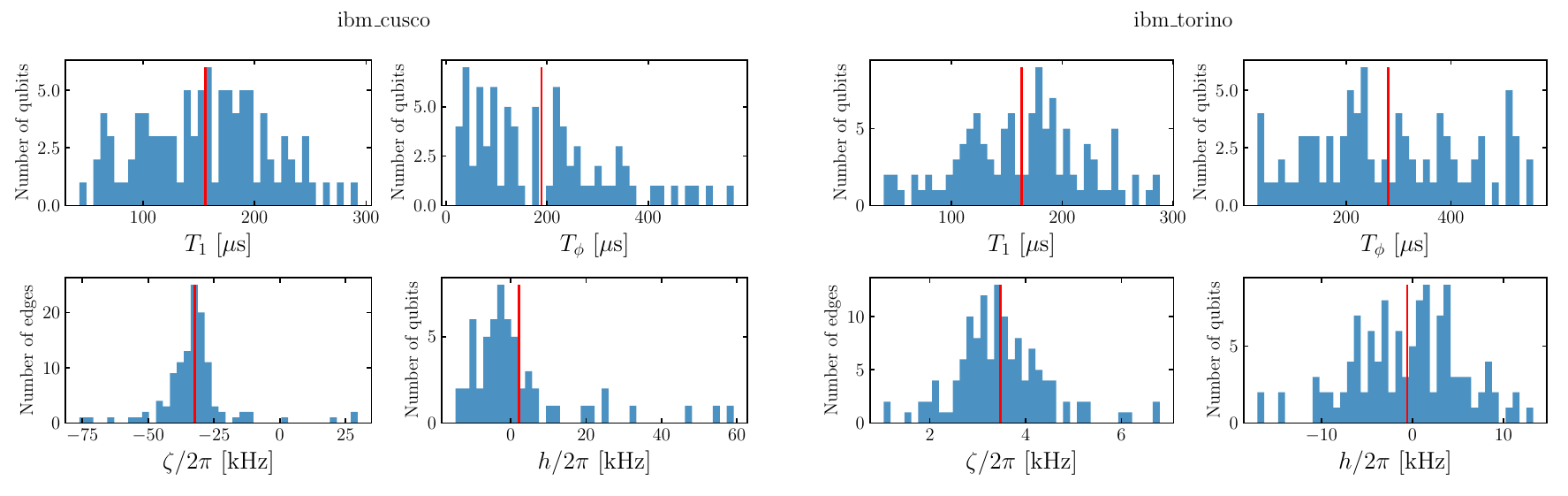}
    \caption{Distributions of the parameters $T_1, T_\phi, h$ and $\zeta$ for deployed IBM Quantum superconducting devices. The red horizontal lines are the mean value of each parameter. In the left panel, an Eagle processor -- \emph{ibm\_cusco}, with 127 qubits. In the right panel, a Heron processor -- \emph{ibm\_torino}, with 133 qubits. Bad qubits that could not be fitted are omitted, and the distribution parameters are summarized in Tab.~\ref{tab:table_exp}.}  
    \label{fig:device_parameter}
\end{figure*}

\section{Experimental distribution of the model parameters in a superconducting-qubit device}\label{sec:experiment}

In this section, we show experimental values for our model parameters in superconducting-qubit device. The values are estimated using characterization experiments \cite{krantz2019quantum, shirizly2024dissipative}, which we run on IBM Quantum processors that are accessible via the cloud \cite{IBMQuantum}. 
As an example, the distributions of the parameters $\left\{T_1, T_\phi, h, \zeta  \right\}$ are shown in \fig{fig:device_parameter} for two types of devices. The summary of parameters estimated in those experiments are given in Tab.~\ref{tab:table_exp}. We note that the detuning error $h$ is distributed relatively symmetrically for \emph{ibm\_torino}, and is somewhat biased to negative values on \emph{ibm\_cusco}, with the sample standard deviation relatively large due to outliers of the distribution. The main difference between the two devices in the context of this work is that the 2Q crosstalks are reduced by one order of magnitude in the Heron processor.

\begin{table}
\begin{tabular}{|c|c|c|}
\hline 
 & Eagle & Heron \tabularnewline
\hline 
\hline 
Device name  & \emph{ibm\_cusco} & \emph{ibm\_torino} \tabularnewline
\hline 
\,Experiment date\, &\, Nov 27th, 2023\, &\, Feb 16th, 2024\, \tabularnewline
\hline 
$T_1$ (stdev) & $ 156\,(55)\,\mu{\rm s}$ & $163\,(56)\,\mu{\rm s}$ \tabularnewline
\hline 
$T_\phi$ (stdev) & $189\,(132)\,\mu{\rm s}$& $280\,(143)\,\mu{\rm s}$ \tabularnewline
\hline 
$h/2\pi$ (stdev) & $2$ $(15)$ kHz  & $-0.6$ $(6.0)$ kHz \tabularnewline
\hline 
$\zeta/2\pi$ (stdev) & $-32$ $(14)$ kHz & $3.5$ $(1.0)$ kHz\tabularnewline
\hline 
\end{tabular}\caption{A summary of the results of characterization experiments estimating device parameters relevant to the current work. See text for details. \label{tab:table_exp}}
\end{table}

\section{Inhomogeneous system dynamics} \label{sec:inhomogeneous}

\begin{figure}
\includegraphics[width=3.3in]{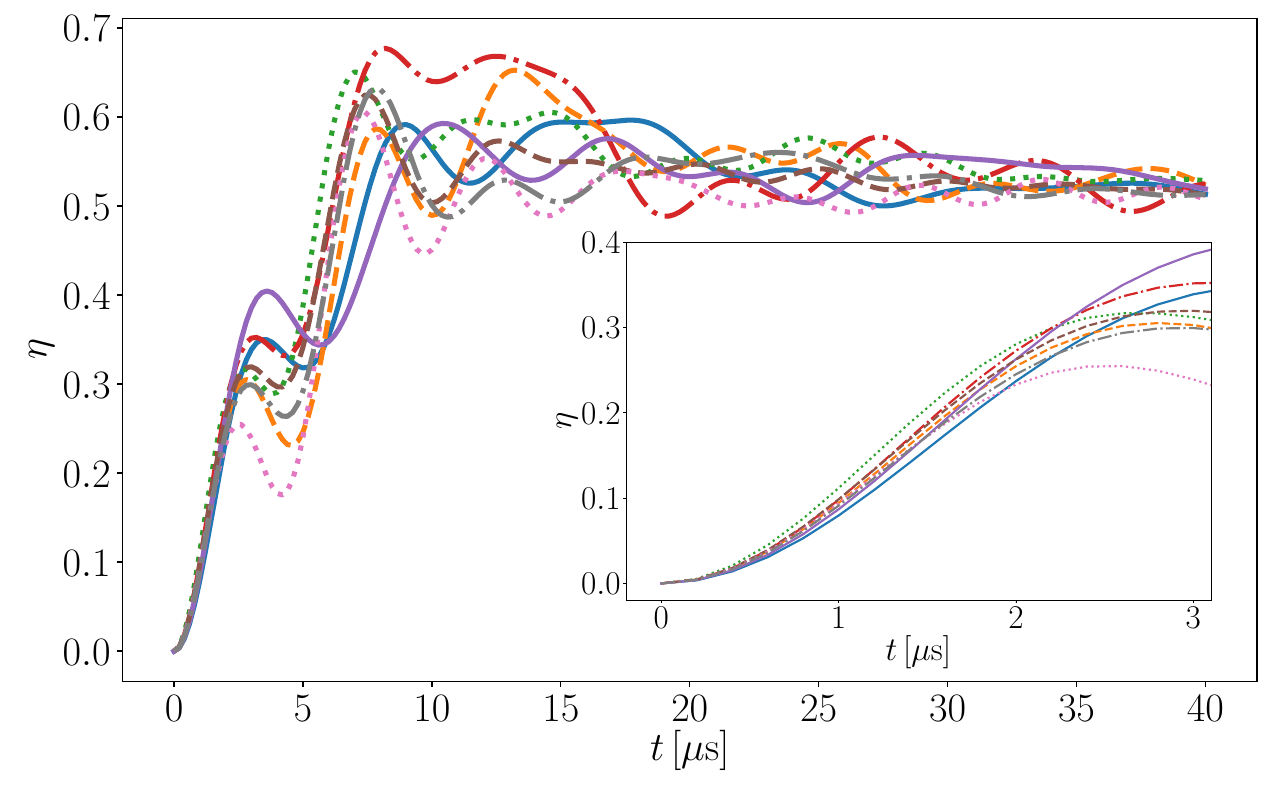}
\caption{The failure probability of eight realizations of the dynamics, each with randomly sampled parameters for the model according to \eqss{eq:dist1}{eq:dist2}. The connectivity (determining the number of nonzero entries of $\zeta_{ij}$) is all-to-all and the initial state is $|0\rangle_L$. Although in every realization the qubits have varying parameters, coherent oscillations in each realization can be seen, with typical timescales determined by the distributions. The inset shows a zoom-in on the initial time, with a visible spread in the growth rate of $\eta$.
} \label{fig:stochastic}
\end{figure}

As a simple approximation of the experimental parameter distributions presented in \app{sec:experiment} for an Eagle device, we sample the parameters of each qubit and each pair for our numerical simulations in the current section from the following heuristic distributions
\be T_{1}\sim {\rm U}[50,150]\,\mu{\rm s}, \quad
T_{\phi}\sim \left( 20 + {\rm Exp}[50]\right)\,\mu{\rm s},\label{eq:dist1}\ee
and
\be \frac{h}{2\pi}\sim \mathcal{N}[0,10]\,{\rm kHz}, \qquad
\frac{\zeta}{2\pi}\sim \mathcal{N}[-30,10]\, {\rm kHz},\label{eq:dist2}\ee
where ${\rm U}[\cdot,\cdot]$ indicates the uniform distribution on the given interval, ${\rm Exp}[\beta]$ is the exponential distribution with mean and standard deviation $\beta$, and $\mathcal{N}[\mu,\sigma]$ is the normal distribution with mean $\mu$ and standard deviation $\sigma$.

Figure \ref{fig:stochastic} shows the dynamics of $\eta$ in eight different realizations with the parameters of each qubit sampled from the distributions above. Coherent oscillations in each realization can be seen, and their relative synchronization could be attributed to $\zeta_{ij}$ values having a well defined mean separated from 0. It is beyond the scope of the current work to explore these aspects in more detail, however, the spread among the different realizations that is seen also at relatively short times (in the figure inset) demonstrates an expected sensitivity in success probabilities of error correction to fluctuations of the qubit parameters, in time and space (among different qubits). 

\section{Pseudo-threshold computations}\label{App:pseudo}

\subsection{Pseudo-threshold computations}
We are presenting in Fig. \ref{fig:pseudo comp} an example of the plots from which we extract the pseduo-threshold as the crossing point between the $\eta$ of the logical qubit and the infidelity of the physical qubit.

\begin{figure}
\centering
    \includegraphics[width=0.44\textwidth]{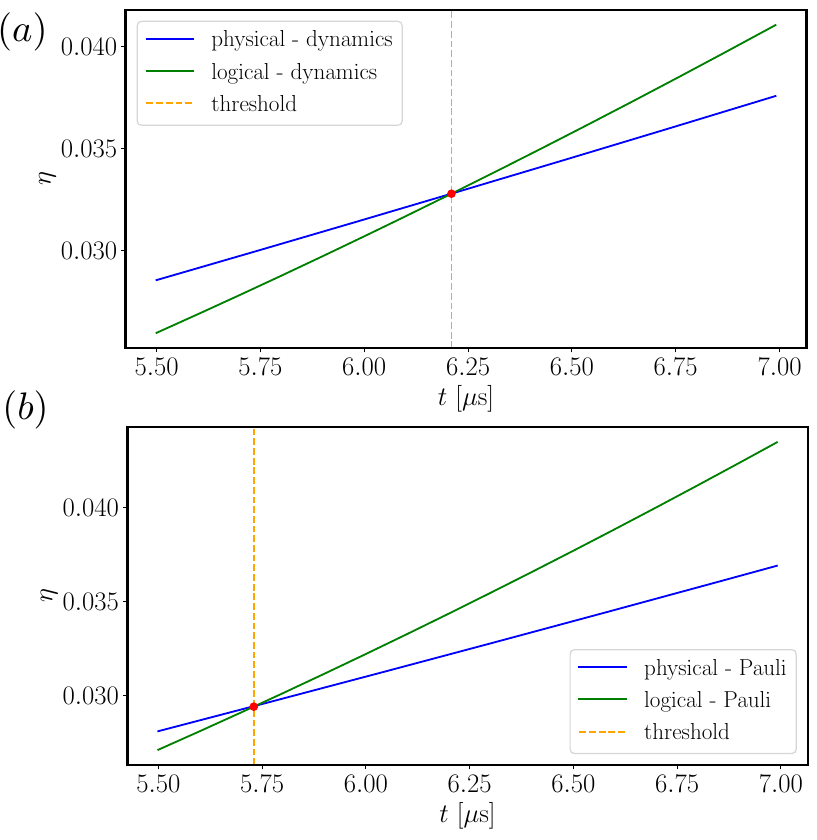}
\caption{\label{fig:pseudo comp} The identification of the pseudo-threshold as the crossing point between the $\eta$ of the logical qubit and the infidelity of the physical qubit, for noise parameters as in Fig. \ref{fig:heron}. (a) for the dynamical noise (b) for the Pauli noise  
}
\end{figure}

\begin{figure}
\centering
\includegraphics[width=0.45\textwidth]{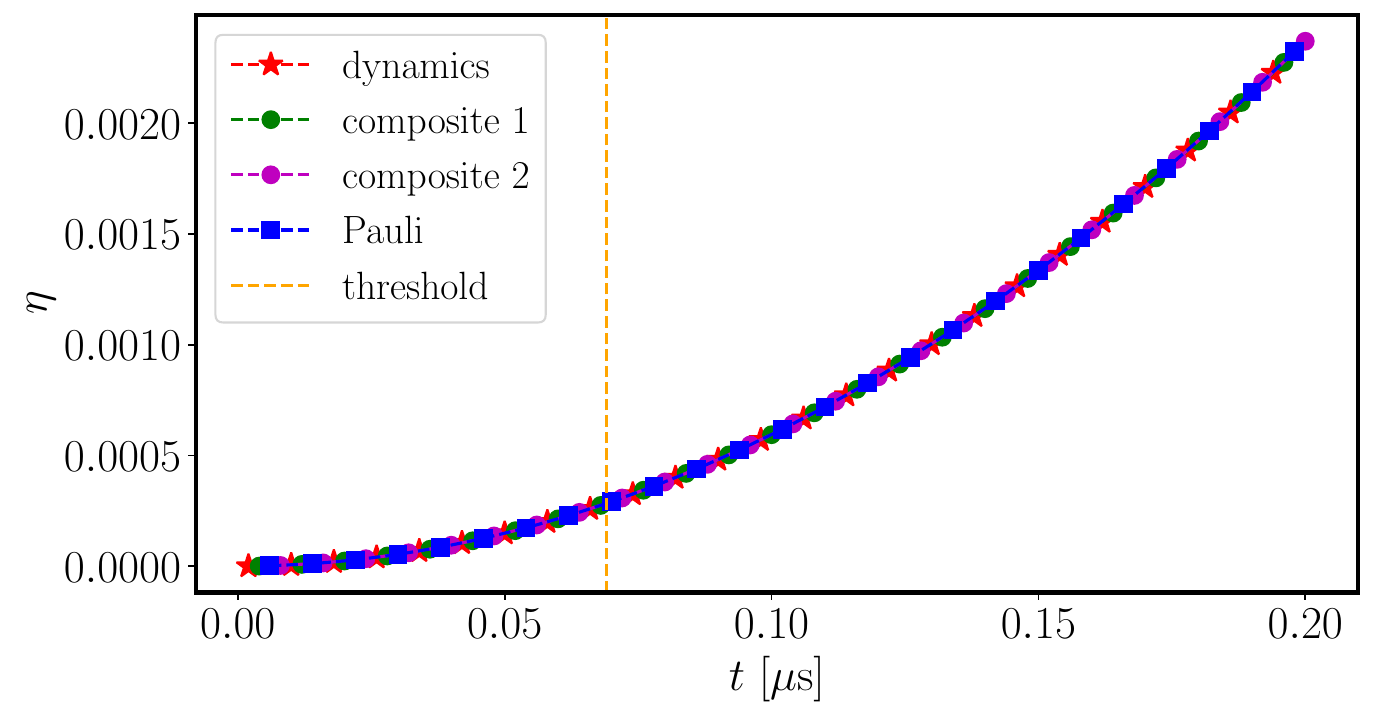}
\caption{\label{fig:pseudo on eagle} The failure probability after error recovery -- $\eta$ of \eq{eq:eta} -- vs.~$t$, the duration of the noise channel, and the noise parameters in \eqss{eq:EagleParams1}{eq:EagleParams2}, with dominant crosstalk noise on an average over the initial states in \eq{eq:states to average}, in the short-time regime, showing that all approximations are identical. The pseudo-threshold for this case is $t=0.069 \mu s$ and we can see that all approximations are identical up to the pseudo-threshold.
}
\end{figure}

\subsection{Pseudo-threshold for other noise parameters}
 When computing the physical error rate, we assumed a single qubit standing idle, and so it does not suffer from two-qubit interaction at all. In a sense this gives an advantage to the physical qubit, since it is oblivious to an important noise source. In practice, when the qubits interact, the physical qubit will also suffer from two-qubit noise, but since we are analyzing only a single logical qubit, we are not dealing with that for now.
And so, in the case of pure ZZ noise, where there is no single qubit noise at all, then the physical qubit does not suffer from any noise, and so there will be no pseudo-threshold. Also, for the cases with eagle devices the threshold will be extremely low, because the dominant noise source is the two qubit interaction.
For completeness, we present in Fig. \ref{fig:pseudo on eagle} the $\eta$ for the eagle parameters as in Fig. \ref{fig:cusco} with the pseudo-threshold.

\subsection{Dynamical noise on a physical qubit}
In all the above, we compared the same noise channel on the physical
qubit and on the logical qubit. Remembering that the physical qubit
is much simpler than the logical qubit, we consider also the case
where simulating the dynamical simulation of the noise is possible
for the physical qubit, and not for the logical qubit. In such a case it will be beneficial to identify how much of the discrepancy in the pseudo-threshold is caused by the approximation of the noise on the logical qubit, and how much is caused by the approximation of the noise on the physical qubit, since the latter can be eliminated by exact simulation of the noise. When we simulate dynamical noise on the physical qubit, and approximated noise on the logical one, 
the pseudo-thresholds of the composite approximations do not
change, but the Pauli noise pseudo-threshold becomes $5.84\,\mu$s.
This means that only a small part of the mismatch was caused by the
inaccuracy of the Pauli approximation on the physical qubit.

\section{Trotter approximation}\label{App:trotter}
A well-known method for solving the time evolution of the density matrix, which doesn't require a full matrix exponential of the Lindbladian, is based on the Trotter approximation. Given a Hamiltonian decomposed as a sum of individual terms, \be H = \sum_i H_i, \ee if all the terms $H_i$ commute in pairs then the time evolution propagator is simply the product of all individual propagators, $U_i(t) = e^{-iH_i t}$. This isn't the general case, however for $t\ll 1$ and noncommuting operators $\hat{A}$ and $\hat{B}$, the propagator can be approximate as 
\be e^{-it\left(\hat{A}+\hat{B}\right)} = e^{-it\hat{A}}e^{-it\hat{B}} + O(t^2)\ee
The trotter approximation is slicing the total time evolution as 
\be e^{-i H t} = \left(\prod_i e^{-i H_i t/n} \right)^n \ee
with $n$ the number of trotter steps.  

Similarly, the Lindbladian evolution can be approximated using a Trotterization of the different noncommuting channels (\apporsm{App:commuting_channels}). This approach can be fruitful since we usually have some information about the individual terms and the channels that are generated. The simplest case would be obtained by composing the channels one after the other, which is equivalent to an approximation using $n=1$ Trotter steps.

\section{Commutation relations between channels}\label{App:commuting_channels}

The distinction between the commutation of observables or operators and the commutation
of quantum channels, describing the evolution of a quantum system,
is important. 
The Pauli matrices $\sigma^{x}$ and $\sigma^{z}$ do not commute in the observable
sense, and this is reflected in their nonvanishing commutator $\left[\sigma^{z},\sigma^{x}\right]=2i\sigma^{y}$.

However, when considering the commutation of channels, the focus shifts
to the resulting state after successive application of the channels
and whether it is affected by the ordering of the channels. Pauli
matrices, serving as unitary matrices, define unitary channels $\mathcal{E}_{a}\left(\rho\right)=\sigma_{a}\rho\sigma_{a}$.
Evaluating $\mathcal{E}_{x}\left(\mathcal{E}_{z}\left(\rho\right)\right)$
and $\mathcal{E}_{z}\left(\mathcal{E}_{x}\left(\rho\right)\right)$
yields the same result ($\sigma^{y}\rho\sigma^{y}$), indicating that the Pauli channels $\mathcal{E}_{x}$ and $\mathcal{E}_{z}$ do
commute. This extends to demonstrate that any Pauli channel commutes
with any other Pauli channel.

In general, the non-commutation of channels can be checked by examining their Kraus representations. Two channels $\mathcal{E}_A\left(\rho\right)=\sum_{a}A_{a}\rho A_{a}^{\dagger}$
and $\mathcal{E}_B\left(\rho\right)=\sum_{b}B_{b}\rho B_{b}^{\dagger}$
commute if and only if the following condition is met 
\begin{equation}\label{Eq:channel_commute}
\sum_{a,b} \overline{A_{a}B_{b}} \otimes A_{a}B_{b} = \sum_{a,b} \overline{B_{b}A_{a}} \otimes B_{b}A_{a}.
\end{equation}
Using \eq{Eq:channel_commute}, we can verify that amplitude damping channel
and phase damping channel commutes. 
In the case of, $\mathcal{U}^{zz}_{\text{2Q}}$
and $\mathcal{K}_{\text{damping}}$, this condition is not satisfied. Consequently,
they do not commute as quantum channels.

The condition in \eq{Eq:channel_commute} is simply shown by using the column-stacking vectorized form  \cite{wood2011tensor} where a matrix \be A = \sum_{i,j}  A_{i,j}\ket{i}\bra{j},\ee is vectorized as \be \dket{A} = A_{i,j}\ket{j} \otimes\ket{i}. \ee 
In this notation, a product of three matrices $ABC$ is given by \be \dket{ABC} = \left( C^T \otimes A \right) \dket{B}. \ee Therefore a channel $\mathcal{E}_A\left(\rho\right)=\sum_{a}A_{a}\rho A_{a}^{\dagger}$ can be written as \be \dket{\mathcal{E}_A\left(\rho\right)} = \hat{\mathcal{E}}_A \dket{\rho} = \sum_a   
\left( \overline{A}_a \otimes A_a \right) \dket{\rho}, \ee
thus two channels $\mathcal{E}_A$ and $\mathcal{E}_B$ commute if and only if 
\be
\hat{\mathcal{E}}_B \circ \hat{\mathcal{E}}_A = \hat{\mathcal{E}}_A \circ \hat{\mathcal{E}}_B,
\ee
which is exactly the condition in \eq{Eq:channel_commute}.

\section{Collapse of the noise channel due to syndrome measurement}\label{sec:channel_collapse}

There is a prevalent belief that a general noise channel
transforms into a Pauli channel post-syndrome measurement. While this
assertion is not entirely accurate, it does hold some validity. The
syndrome measurement indeed alters the noise channel, rendering it
more akin to a Pauli channel by causing the elimination of numerous
non-diagonal terms, although not all of them. In the following analysis,
we will provide a straightforward examination of the terms that vanish
and those that persist in this transformation.

Every quantum measurement can be characterized by a set of measurement
operators $\left\{ M_{k}\right\} $ with $\sum_{k}M_{k}^{\dagger}M_{k}=I$,
where the probability $p_{k}=\text{Tr}\left[M_{k}\rho M_{k}^{\dagger}\right]$
corresponds to measuring the value associated with $M_{k}$. Following
the measurement and averaging over possible outcomes, the state $\rho$
collapses into the state $\sum_{k}M_{k}\rho M_{k}^{\dagger}$ \cite{lidar2019lecture}.
For the measurement of a stabilizer $S$ we can define the measurement
operators as $M_{+}=\frac{1}{2}\left(I+S\right)$ and $M_{-}=\frac{1}{2}\left(I-S\right)$.
When applied to the noisy state $\mathcal{E}\left(\rho\right)=
\sum_{k}E_{k}\rho E_{k}^{\dagger}=\sum_{k}\sum_{\sigma\in\mathcal{P}}\sum_{\nu\in\mathcal{P}}a_{k\sigma}a_{k\nu}^{*}\sigma\rho\nu$,
we obtain
\begin{align}
& M_{\pm}\mathcal{E}\left(\rho\right)M_{\pm} \nonumber \\
& =\frac{1}{2}\left(I\pm S\right)\sum_{k}\sum_{\sigma\in\mathcal{P}}\sum_{\nu\in\mathcal{P}}a_{k\sigma}a_{k\nu}^{*}\sigma\rho\nu\frac{1}{2}\left(I\pm S\right)\nonumber \\
 & =\frac{1}{4}\sum_{k}\sum_{\sigma\in\mathcal{P}}\sum_{\nu\in\mathcal{P}}a_{k\sigma}a_{k\nu}^{*}\left(\sigma\rho\nu\pm S\sigma\rho\nu\pm\sigma\rho\nu S+S\sigma\rho\nu S\right).
\end{align}

Since the stabilizer $S$ is a Pauli operator, it either commutes or anti-commutes with every Pauli $\sigma$. Consequently, we define 
\be d_{S,\sigma}=\begin{cases}
0 & \left[S,\sigma\right]=0\\
1 & \left\{ S,\sigma\right\} =0
\end{cases}.\ee
Additionally, starting with an initial state $\rho=|\psi\rangle\langle\psi|$
that is a logical code state, we have $S|\psi\rangle=|\psi\rangle$. These
two conditions lead to
\begin{align}
& M_{\pm}\mathcal{E}\left(\rho\right)M_{\pm} \nonumber \\
 & =\frac{1}{4}\sum_{\sigma\in\mathcal{P}}\sum_{\nu\in\mathcal{P}}\left(\sum_{k}a_{k\sigma}a_{k\nu}^{*}\right)f\left(S,\sigma,\nu\right)\sigma\rho\nu,
\end{align}
having defined \be f\left(S,\sigma,\nu\right)=1\pm\left(-1\right)^{d_{S,\sigma}}\pm\left(-1\right)^{d_{S,\nu}}+\left(-1\right)^{d_{S,\sigma}+d_{S,\nu}}.\ee
When $d_{S,\sigma}\neq d_{S,\nu}$ then $f\left(S,\sigma,\nu\right)=0$.
Indeed, only terms sharing the same commutation relation with the
stabilizer will persist in the sum. Extending this analysis across
all stabilizers, we find that surviving terms post-stabilizer measurement
are of the form $\sigma\rho\nu$ where $\sigma$ and $\nu$ exhibit
identical commutation relations with all stabilizers. While a majority
of terms vanish, the diagonal terms $\sigma\rho\sigma$ all survive.
However, there are additional surviving terms, including those where
$\sigma$ and $\nu$ differ by a stabilizer or a logical operator.
Thus, the Pauli approximation proves to be a generally effective approximation,
although its success is contingent on the characteristics of the original
noise channel. As demonstrated in \cite{gutierrez2016errors} the
Pauli approximation excels for incoherent noise but falls short for
coherent noise.

\section{Pauli approximation methods}\label{sec:pauli methods}

It is important to note that the most precise way to perform the Pauli approximation is on the full 5-qubit quantum channel. That is, representing the full noise, including all 2Q terms and all 1Q terms as one channel in the form of \eq{eq:general channel} and performing the approximation as in \eq{eq:Pauli} on the full channel. But in order to perform this, we need a representation of the full quantum channel.
Obtaining this representation is a non trivial goal, and becomes more complicated and time consuming as the code becomes larger. 
Moreover, the approximation we will obtain will have many terms, some with very high weight, i.e. a large portion of the qubits with a non-trivial Pauli. This makes the approximation complicated to implement, not much simpler than implementing the full channel.
And so we will essentially be settling for the much less accurate Pauli approximation instead of applying the accurate full channel, with hardly any gain in computational resources.

And so instead of doing the Pauli approximation on the full noise channel, we will separate the full quantum channel into smaller groups, and perform the Pauli approximation separately on each such group.
Before doing so we will like to give an example of how the Pauli approximation of a full channel differs from the Pauli approximation of parts of the channel separately.
As an example for this effect, let us consider the simple 1Q unitary channel $\mathcal{E}_{U}\left(\rho\right)=U\rho U^{\dagger}$
given by 
$U=\left(\begin{matrix}1 & 0\\
0 & e^{i\varphi}
\end{matrix}\right)$.
If we apply a Pauli approximation on the full channel we get
\be
\mathcal{E}_{U}^P\left(\rho\right)=\cos^{2}\left(\frac{\varphi}{2}\right)\rho+\sin^{2}\left(\frac{\varphi}{2}\right)Z\rho Z.
\ee
Now, if instead we would separate this into two applications of the
channel $\mathcal{E}_{1/2}\left(\rho\right)=U_{1/2}\rho U_{1/2}^{\dagger}$
where $U=\left(\begin{matrix}1 & 0\\
0 & e^{i\varphi/2}
\end{matrix}\right)$, then $\mathcal{E}_{1/2}\left(\mathcal{E}_{1/2}\left(\rho\right)\right)=\mathcal{E}_{U}\left(\rho\right)$.
But if we perform the Pauli approximation on each part separately,
we get
\be
\mathcal{E}_{\frac{1}{2}}^{P}\left(\rho\right)=\cos^{2}\left(\frac{\varphi}{4}\right)\rho+\sin^{2}\left(\frac{\varphi}{4}\right)Z\rho Z,
\ee
and so 
\be
\mathcal{E}_{\frac{1}{2}}^{P}\left(\mathcal{E}_{\frac{1}{2}}^{P}\left(\rho\right)\right)=\frac{1}{4}\left(\cos\varphi+3\right)\rho+\frac{1}{2}\sin^{2}\frac{\varphi}{2}Z\rho Z\neq\mathcal{E}_{U}^{P}\left(\rho\right).
\ee

Establishing that the accuracy is lower when the approximation is done on separate groups, we are willing to sacrifice this accuracy in order to gain simplicity and speed of implementation.
The most intuitive way to do this approximation, and the one we implemented in the paper, is applying the Pauli approximation on the separate channels depicted in \fig{fig:circuit sketchs}(b).
Remembering that the composite channel consists of 2Q terms and 1Q terms, we group all 1Q channels acting on a specific qubit into a unified 1Q channel, and then apply a Pauli approximation on this channel. Similarly, we group all 2Q channels acting on a specific pair of connected qubits, and then apply the Pauli approximation on this channel.

Another method that may seem to have certain advantages is to group all the 1Q channels together with the 2Q channels and apply the Pauli approximation on these unified 2Q channels. This can be done, for example, by separating every 1Q channel on a qubit which is connected to $n$ other qubits into $n$ identical channels, where applying them consecutively gives the original 1Q channel. Next each one of the $n$ pieces is appended to each of the 2Q channels, and then we only need to apply the Pauli approximation on the 2Q channels. 
This may seem to be beneficial since we apply the Pauli approximation less times, but in practice does not necessarily give a better approximation. This is due to the errors we get from approximating the 1Q channel on a specific qubit $n$ times, instead of once. This introduces errors that overshadow, in many cases, any advantage that this method may seem to have.
And so we decided to approximate the 1Q and 2Q channels separately.

\section{2Q XY coupling noise}\label{sec:XY}

\begin{figure}
\centering
    \includegraphics[width=0.48\textwidth]{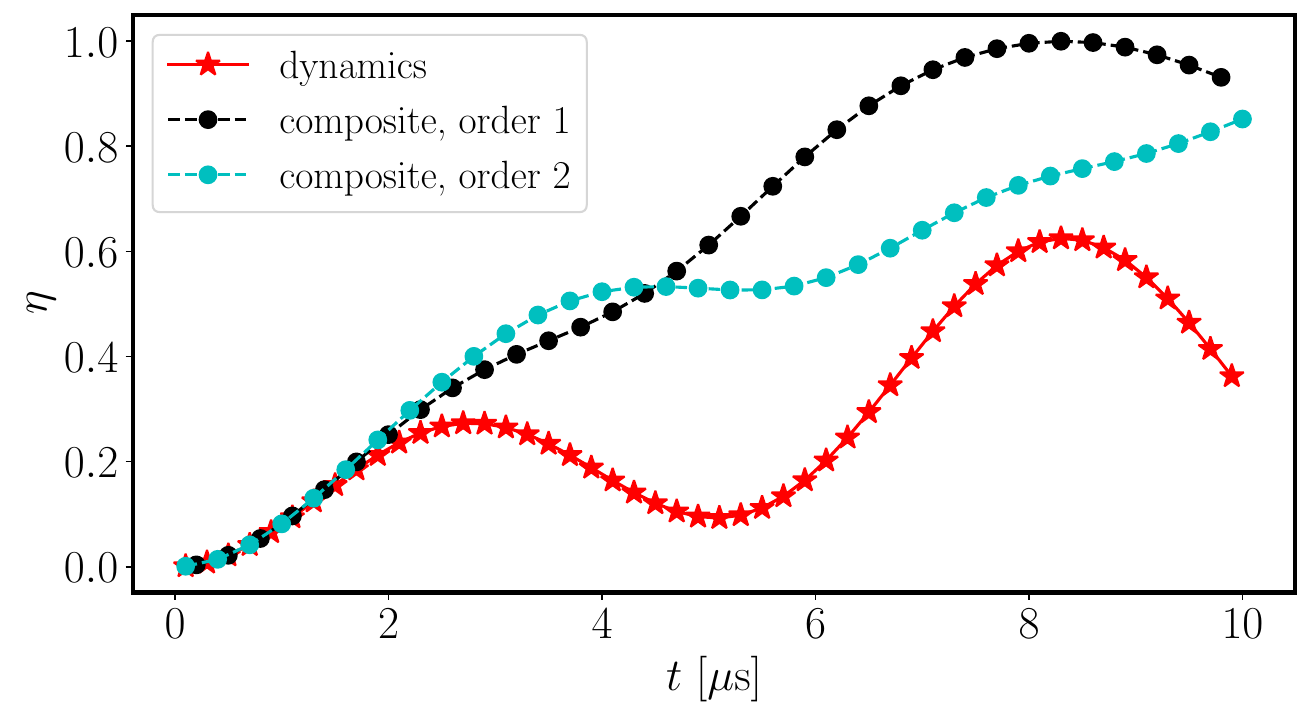}
\caption{\label{fig:jxy} $\eta$ vs.~$t$ with the initial state $|0\rangle_L$ and the being only XY coupling with $J=-2\pi\times 30\,$kHz.
We can see here qualitative inaccuracies in the results obtained by the composite-channel approximation.
}
\end{figure}

In this section, we briefly consider an XY term (also called ``flip-flop'' or an exchange coupling), instead of the ZZ coupling considered in the rest of the paper. Hence, we take the Hamiltonian to be
\begin{equation}
\mathcal{H}/\hbar =\sum_{i}H_{i}+\sum_{\left\langle i,j\right\rangle }V^{xy}_{ij},\label{eq:hamiltonian_xy}
\end{equation} 
with $H_i$ defined in \eq{Eq:1Q_ham},
and
\be
V^{xy}_{ij}= \frac{1}{2}J_{ij}  \left(\sigma^x_{i} \sigma^x_{j} + \sigma^y_{i} \sigma^y_{j}\right).
\end{equation}

The $V^{xy}$ term gives rise to the channel $\mathcal{U}^{xy}_{\text{2Q}}$ in the composite-channel approximation,
\be \mathcal{U}^{xy}_{\text{2Q}}\left(\rho\right)=U_{3}\rho U_{3}^{\dagger},\label{eq:K2Qxy}\ee
with 
\begin{equation}
U_{3}=e^{-i \left( V^{xy}_{12}\right)t}=\left(\begin{matrix}1 & 0 & 0 & 0\\
0 & \cos Jt  & -i\sin Jt  & 0\\
0 & -i\sin Jt & \cos Jt & 0\\
0 & 0 & 0 & 1
\end{matrix}\right).
\end{equation}

The channel above doesn't commute even with itself when there is only a single common qubit, e.g., when applying the channel on qubits 1 and 2 and on qubits 2 and 3, the order of these applications makes a difference.
With 5 physical qubits, there are many different orderings one can choose.
The results of two different orderings are presented in \fig{fig:jxy}. In both cases we begin by applying all the 1Q channels. We label the physical qubits $1-5$. We partition the 2Q gates into groups where group $n$ consists of all 2Q channels which affect qubit number $n$ but do not affect qubit $i$ for any $i<n$. Inside each group, the channels are applied by the natural order on the second qubit affected by the channel.
Order 1 in Figure \ref{fig:jxy} is applying the channel groups by natural order: group 1, 2, 3, 4.
Order 2 is applying the channel groups by the following order: 2, 3, 1, 4.

Now, while the commutator in this case is of second order in the noise parameters as in the case of $\mathcal{U}^{zz}_{\text{2Q}}$ with the 1Q damping, for the studied range of noise parameters, the $\mathcal{U}^{xy}_{\text{2Q}}$ commutators can be significantly larger since $J\gg 1/T_1$ (a more detailed analysis can be found in \app{sec:composite big t1}), and so we would expect a more significant error of the composite-channel approximation in this case.
Indeed, we can see in \fig{fig:jxy} that for $t\gtrsim 2.5\,\mu$s the composite-channel approximation no longer holds, not even qualitatively.

\section{Failure of the composite-channel approximation}
\label{sec:composite big t1}

\subsection{Failure for XY coupling}

To see how the approximation of composing the channels of each XY interaction breaks, we can focus on 3 qubits in a chain with the Hamiltonian 
\begin{align}
& H/\hbar = V^{xy}_{1,2} + V^{xy}_{2,3} \nonumber \\
& = \frac{1}{2}J\left(\sigma^x_1 \sigma^x_2 +\sigma^y_1 \sigma^y_2\right) + \frac{1}{2}J\left(\sigma^x_2 \sigma^x_3 +\sigma^y_2 \sigma^y_3\right), 
\end{align}
to clearly see how the composite propagation 
\be e^{-it\left(V^{xy}_{1,2} + V^{xy}_{2,3}\right)} = e^{-it V^{xy}_{1,2}}e^{-it V^{xy}_{2,3}} + O(t^2)\ee
stops being valid.
The XY interaction conserve the number of excitation, therefore the Hamiltonian is block diagonal with respect to the states $\left\{\ket{000}\right\}$,$\left\{\ket{100},\ket{010},\ket{001}\right\}$,$\left\{\ket{011},\ket{101},\ket{110}\right\}$ and $\left\{\ket{111}\right\}$. 

With respect to the block $\left\{\ket{100},\ket{010},\ket{001}\right\}$, the time evolution of  
the individual terms are 
\be
 e^{-it V^{xy}_{1,2}} =\left(\begin{matrix} \cos{\left(J t \right)} & - i \sin{\left(J t \right)} & 0\\- i \sin{\left(J t \right)} & \cos{\left(J t \right)} & 0\\0 & 0 & 1\end{matrix}\right)
\ee
\be
 e^{-it V^{xy}_{2,3}} = \left(\begin{matrix}1 & 0 & 0\\0 & \cos{\left(J t \right)} & - i \sin{\left(J t \right)}\\0 & - i \sin{\left(J t \right)} & 1 \cos{\left(J t \right)}\end{matrix}\right) .
\ee
Then, composing the propagators gives

\begin{align}
& e^{-it V^{xy}_{1,2}}e^{-it V^{xy}_{2,3}} = \nonumber \\
&\left(\begin{matrix} \cos{\left(J t \right)} & - 0.5 i \sin{\left(2 J t \right)} & - \sin^{2}{\left(J t \right)}\\- i \sin{\left(J t \right)} & \cos^{2}{\left(J t \right)} & - 0.5 i \sin{\left(2 J t \right)}\\0 & - i \sin{\left(J t \right)} & \cos{\left(J t \right)}\end{matrix}\right),
\label{Eq:Jsep}
\end{align}
in comparison to the full Hamiltonian, which is

\begin{align}
& e^{-it\left(V^{xy}_{1,2} + V^{xy}_{2,3}\right)}  = \nonumber \\ 
& \left(\begin{matrix}\frac{1}{2} \cos{\left(\sqrt{2} J t \right)} + \frac{1}{2} & - \frac{1}{\sqrt{2}} i \sin{\left(\sqrt{2} J t \right)} & \frac{1}{2} \cos{\left(\sqrt{2} J t \right)} - \frac{1}{2}\\- \frac{1}{\sqrt{2}} i \sin{\left(\sqrt{2} J t \right)} & \cos{\left(\sqrt{2} J t \right)} & - \frac{1}{\sqrt{2}} i \sin{\left(\sqrt{2} J t \right)}\\\frac{1}{2} \cos{\left(\sqrt{2} J t \right)} - \frac{1}{2} & - \frac{1}{\sqrt{2}} i \sin{\left(\sqrt{2} J t \right)} & \frac{1}{2} \cos{\left(\sqrt{2} J t \right)} + \frac{1}{2}\end{matrix}\right).
\label{Eq:Jboth}
\end{align}

We note that \eq{Eq:Jboth} is consistent with \eq{Eq:Jsep} up to linear order in $Jt$, however the frequency of the sines and cosine differ, causing the dynamics to deviate.

\subsection{Failure for ZZ crosstalk and large damping noise}\label{sec:T_1noncummutativity}

The composite-channel approximation fails to properly incorporate the non-trivial commutation relations between the noise terms. In \fig{fig:jxy} we see that the composite relation fails at $t\simeq 2.5 \mu s$ in the presence of $\mathcal{U}_{\text{2Q}}^{xy}$ noise. In this case, $J=2\pi\times30$ kHz and so the commutator is proportional to $(J t)^2\simeq 0.22$, and is indeed nearing one. 
On the other hand, in \fig{fig:cusco} where the non commuting terms are $\mathcal{U}_{2Q}^{z}$ and the damping noise, the composite-channel approximation holds very well.
In this case, the noise parameters, and specifically the damping parameters are relatively small. For $\zeta=-2\pi\times 30\,$kHz and $T_1=150 \mu s$, the commutator for $t= 2.5 \mu s$  is proportional to $ |\zeta |t \frac{t}{T_1}\approx 0.008$. When analyzing the non realistic case, where $T_1=5 \mu s$ giving rise to $ |\zeta| t \frac{t}{T_1}\approx 0.24$, then as can be seen in \fig{fig:tiny t1}, the composite-channel approximation differs from the dynamical simulation more significantly than for smaller $|\zeta|$ as in \fig{fig:cusco}, but the deviation is not qualitative as in \fig{fig:jxy}.

\begin{figure}
\centering
    \includegraphics[width=0.48\textwidth]{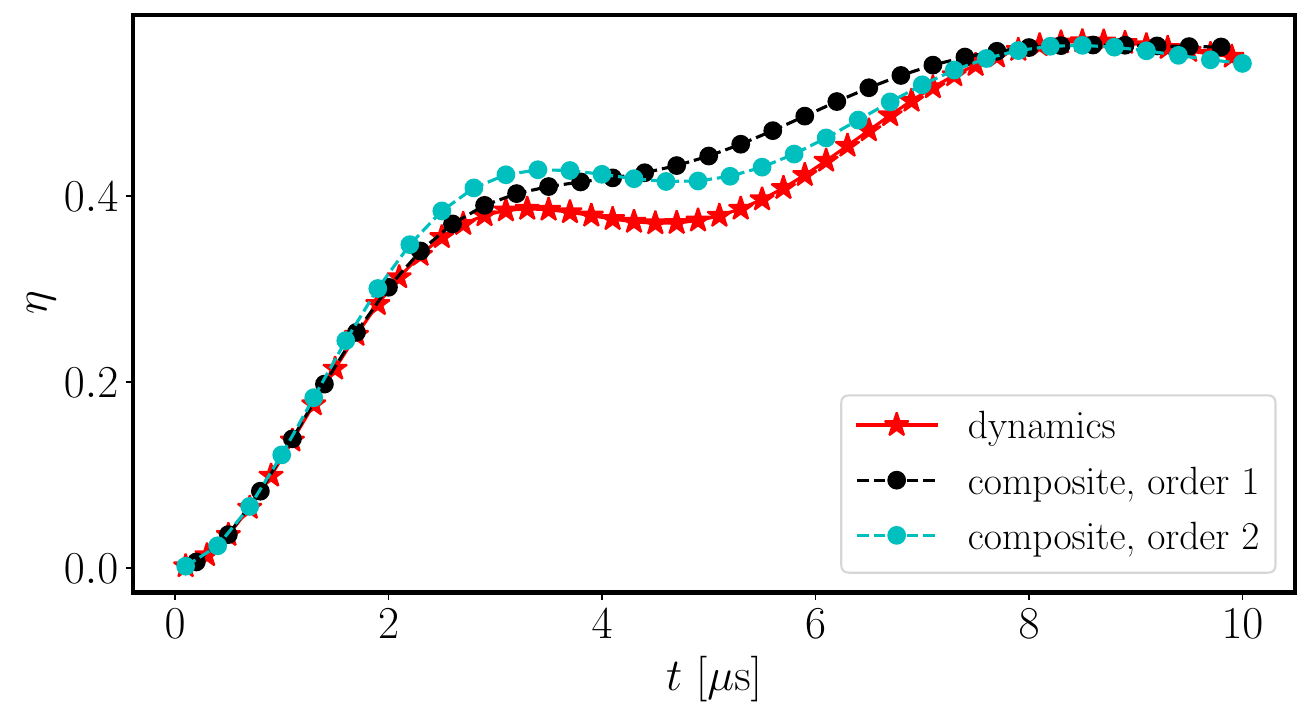}
\caption{\label{fig:tiny t1}Plot of $\eta$ as a measure of the infidelity of the noisy state after correction. The x axis is the time $t$ the noise channel acted on the state. The initial state is $|0\rangle_L$. The noise parameters are $T_1=5 \mu$s, $T_2=10 \mu$s, $\zeta=-2\pi\times 30$ kHz. The x axis up to $t=10 \mu$s.
There is no significant qualitative error in the performance of the composite-channel approximation, even though the noise parameters are very large.
}
\end{figure}

\section{1Q noise and approximations} \label{app:1QApproximation}

Starting with 1Q noise only, i.e., the single qubit Hamiltonian part of \eq{Eq:1Q_ham} together with the dissipative part of the noise $\mathcal{D}_0$ and $\mathcal{D}_2$, all noise terms commute and constitute single-qubit channels, therefore as discussed in \seq{sec:1Q}, the composite-channel approximation is exact and the Pauli approximation works well.
This can indeed be seen in \fig{fig:1Q noise} for the initial state $|+\rangle_L$ for short and intermediate times.
Other initial states lead to similar results.

\begin{figure}
\centering
    \includegraphics[width=0.48\textwidth]{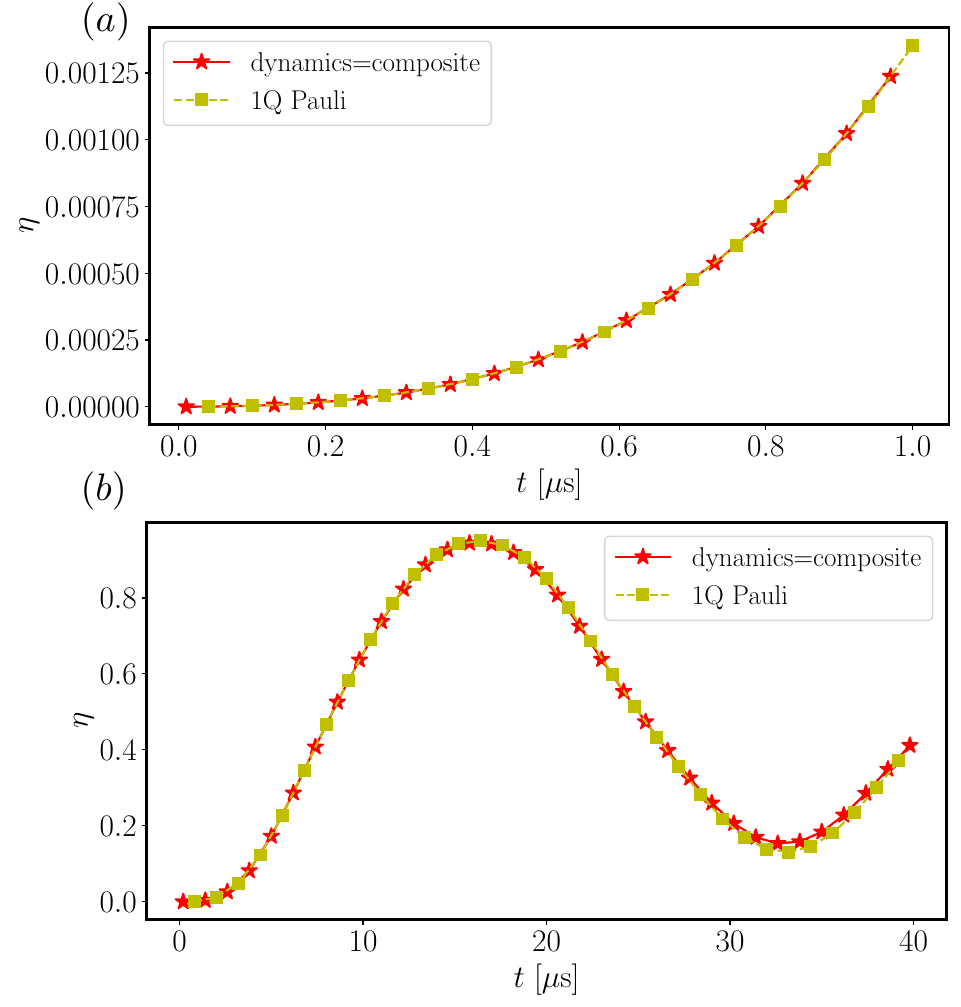}
\caption{\label{fig:1Q noise}The failure probability after error recovery -- $\eta$ defined in \eq{eq:eta} -- vs.~the time $t$ during which the noise channel acted on the state, for (a) short times, and (b) intermediate times. The initial state is $|+\rangle_L$ and the noise parameters are $T_1=150\,\mu$s, $T_2=100\,\mu$s, $h=-2\pi\times 30\,$kHz, and no 2Q noise. 
The figure shows that, as expected, all approximations hold very well in the short time range as well as the intermediate time range.
}
\end{figure}

Our next objective is to acquire a noise channel that operates on
a single qubit, as can be seen in Figure \ref{fig:circuit sketchs}(c). 
This approach will alleviate the necessity of simulating
two-qubit channels. This requires approximating $\mathcal{U}^{zz}_{\text{2Q}}$
by a single qubit channel, 
since the other channels considered here are already 1Q channels.
When a two-qubit unitary acts on a separable
state $|\psi_1\psi_2\rangle$, the effect of this on one of the qubits -- disregarding
the second -- is described by a channel with the Kraus operators $E_{k}=\left\langle e_{k}|U|\psi_2\right\rangle $
where $\left\{ |e_{k}\rangle\right\} $ is a basis for the first qubit.
However,
this approach isn't directly applicable in our case due to the highly
entangled nature of the logical states. 
We will instead try to capture the important traits of $U_2$ in a 1Q channel. Notably, $U_{2}$ is a 2Q unitary which adds a phase $\phi_+=+i \zeta t/2$ when the parity of the qubits is even, and $\phi_-=-i \zeta t/2$ otherwise.
The logical states contain an equal amount of physical $1$s and $0$s and so it seems reasonable to approximate this by a 1Q channel that with probability $1/2$ adds a phase $\phi_+$ to the state $|0\rangle$ and a phase $\phi_-$ to the state $|1\rangle$, and with probability $1/2$ a unitary with these phases reversed. We therefore obtain
\be \tilde{\mathcal{K}}^{(1)} \left(\rho\right)=U_{+}\rho U_{+}^{\dagger}+U_{-}\rho U_{-}^{\dagger},\ee
with
\be U_{+}=\frac{1}{\sqrt{2}}\left(\begin{matrix} e^{-i \zeta t/2} & 0\\
0 & e^{+i \zeta t/2} \end{matrix}\right), \ee 
and
\be U_{-}=\frac{1}{\sqrt{2}}\left(\begin{matrix}e^{+i \zeta t/2} & 0\\
0 & e^{-i \zeta t/2}
\end{matrix}\right). \ee
It is worth noting that we would arrive at the same channel also
if we assumed that the second qubit initially began in the state $\frac{1}{\sqrt{2}}\left(|0\rangle+|1\rangle\right)$ or any other state with symmetric population of 0 and 1. 

A qubit with connectivity $n$ is connected to $n$ other qubits and so is involved in $n$ two qubit channels in the previous level of approximation. So to approximate this by a single qubit channel, the channel $\tilde{\mathcal{K}}^{(1)}$ must be implemented on such a qubit $n$ times, as is depicted in Figure \ref{fig:circuit sketchs}(c). So in total we apply the channel
\begin{align}
\tilde{\mathcal{K}}^{(n)}\left(\rho\right) & = \sum_{k=0}^{n} {n \choose k} U_{+}^{n-k} U_{-}^{k}\rho U_{-}^{\dagger k} U_{+}^{\dagger n-k}\nonumber \\
& = \sum_{k=0}^{n}\frac{1}{2^{n}}{n \choose k} U^P_{1Q}\rho U^{P\dagger}_{1Q},
\end{align}
where $U^P_{1Q}=\left(\begin{matrix}e^{-i \zeta t(n-2k)/2} & 0\\
0 & e^{i \zeta t(n-2k)/2}
\end{matrix}\right)$, on each qubit as the approximation of $\mathcal{U}^{zz}_{\text{2Q}}$. 
This channel commutes with $\mathcal{K}_{1Q}$.

The final approximation is the Pauli approximation obtained from the single-qubit Kraus channel, 
\begin{equation}
\tilde{\mathcal{P}}_{}\left(\rho\right)=c_{0}\rho+c_{x}\sigma^x\rho \sigma^x+c_{y}\sigma^y\rho \sigma^y+c_{z}\sigma^z\rho \sigma^z,
\end{equation}
with
\begin{align}
c_{x} & =c_{y}  =\frac{1}{4}p_{\rm ad}, \\c_{z} & =\frac{1}{2}-\frac{1}{4}p_{\rm ad}-\frac{1}{2}\gamma \sum_{k=0}^{n} \frac{1}{2^{n}}{n \choose k}\cos\left[\left(h_{}+(n-2k)\zeta\right)t\right],
\end{align}
where as before, $c_0=1-c_x-c_y-c_z$ and $\gamma$ is defined in \eq{eq:gamma}.

These approximations failed at estimating the success rate of the error correction for all parameters even for very small time-scales. An example for this is presented in \fig{fig:heron with 1q approx}.

\begin{figure}
\centering
    \includegraphics[width=0.48\textwidth]{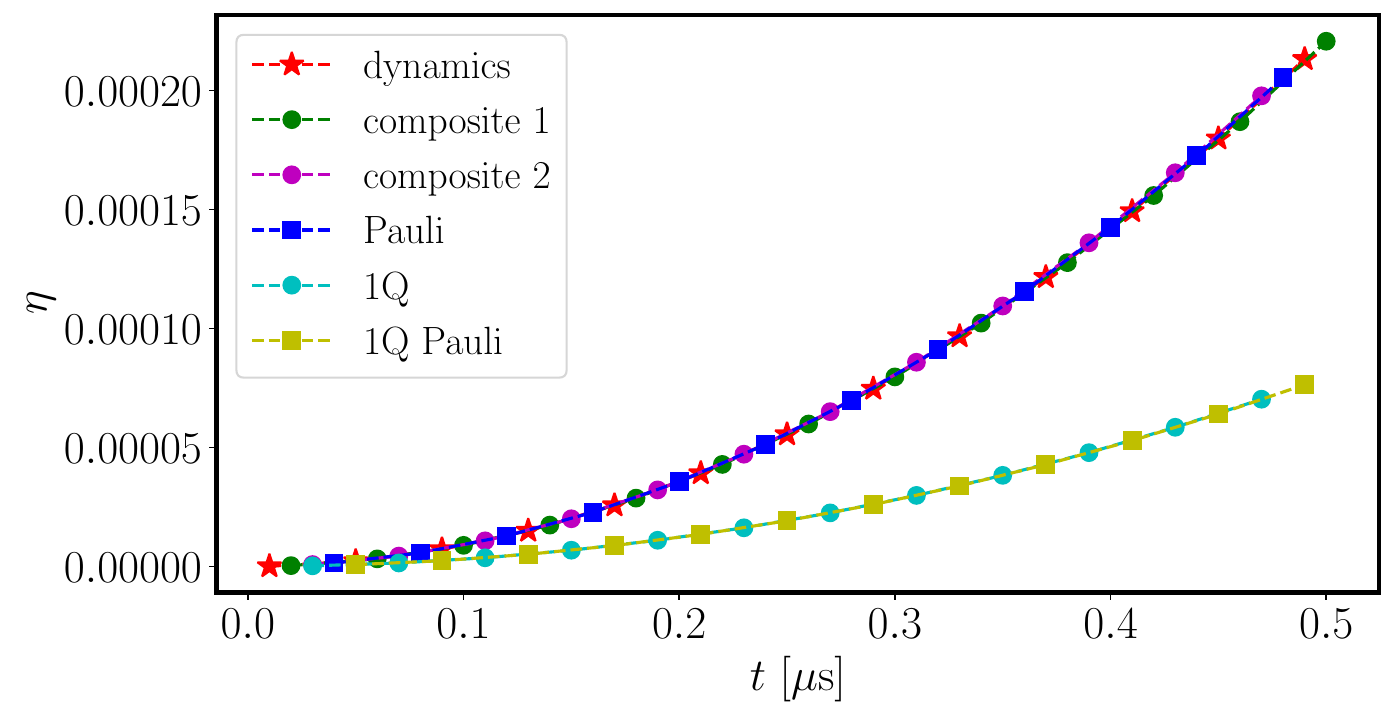}
\caption{\label{fig:heron with 1q approx} The failure probability $\eta$ vs.~$t$ shown for the average over initial states in \eq{eq:states to average}, with the noise parameters in \eqss{eq:HeronParams1}{eq:HeronParams2}. Even for very short time scales the 1Q approximations significantly under-estimate the error correction success rate.
}
\end{figure}

\begin{table}
\begin{tabular}{|c|c|c|c|}
\hline 
Syndrome & Standard & FC decoder & Ring decoder\tabularnewline
\hline 
\hline 
$0000$ & $I$ & $I$ & $I$\tabularnewline
\hline 
$0001$ & $X_{0}$ & $Z_{1}Z_{4}$ & $X_{0}$\tabularnewline
\hline 
$0010$ & $Z_{2}$ & $Z_{2}$ & $Z_{2}$\tabularnewline
\hline 
$0011$ & $X_{4}$ & $Z_{0}Z_{3}$ & $X_{4}$\tabularnewline
\hline 
$0100$ & $Z_{4}$ & $Z_{4}$ & $Z_{4}$\tabularnewline
\hline 
$0101$ & $Z_{1}$ & $Z_{1}$ & $Z_{1}$\tabularnewline
\hline 
$0110$ & $X_{3}$ & $Z_{2}Z_{4}$ & $X_{3}$\tabularnewline
\hline 
$0111$ & $Y_{4}$ & $Z_{1}Z_{2}$ & $Z_{1}Z_{2}$\tabularnewline
\hline 
$1000$ & $X_{1}$ & $Z_{0}Z_{2}$ & $X_{1}$\tabularnewline
\hline 
$1001$ & $Z_{3}$ & $Z_{3}$ & $Z_{3}$\tabularnewline
\hline 
$1010$ & $Z_{0}$ & $Z_{0}$ & $Z_{0}$\tabularnewline
\hline 
$1011$ & $Y_{0}$ & $Z_{2}Z_{3}$ & $Z_{2}Z_{3}$\tabularnewline
\hline 
$1100$ & $X_{2}$ & $Z_{1}Z_{3}$ & $X_{2}$\tabularnewline
\hline 
$1101$ & $Y_{1}$ & $Z_{3}Z_{4}$ & $Z_{3}Z_{4}$\tabularnewline
\hline 
$1110$ & $Y_{2}$ & $Z_{0}Z_{4}$ & $Z_{0}Z_{4}$\tabularnewline
\hline 
$1111$ & $Y_{3}$ & $Z_{0}Z_{1}$ & $Z_{0}Z_{1}$\tabularnewline
\hline 
\end{tabular}\caption{\label{tab:mod-dec}The recovery operator dictated by the different
decoders. Each recovery process is characterized by its action on the qubits it affects non-trivially, for example, $Z_1 Z_3=Z_1 I_2 Z_3 I_4 I_5$.}
\end{table}

\begin{figure}
\centering
    \includegraphics[width=0.48\textwidth]{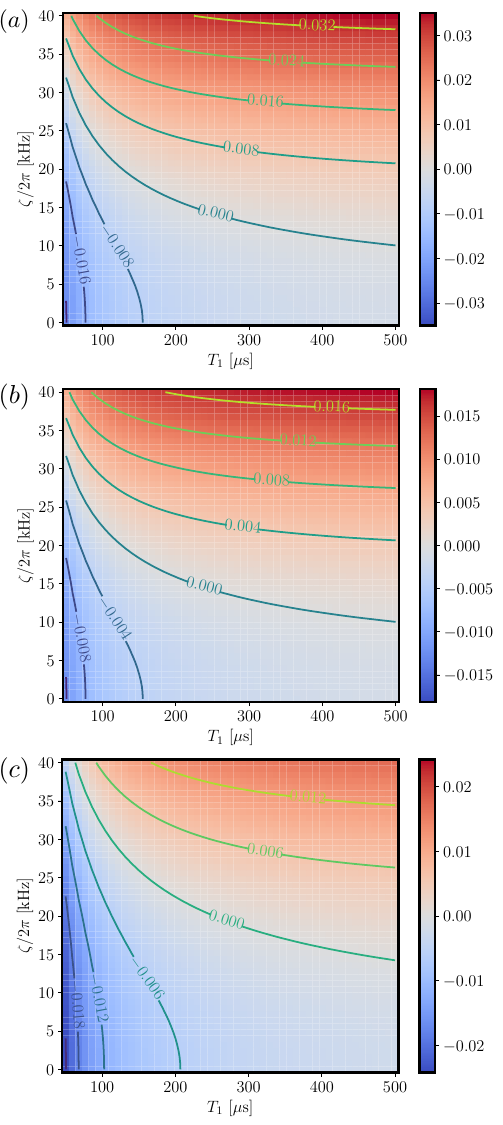}
\caption{\label{fig:modified decoder all}The difference between the logical error when the standard decoder was applied vs.~a modified decoder. $h$ is set to $0$, and $T_2=2T_1$. The areas colored red are parameter values for which the modified decoder is beneficial, while the standard decoder is beneficial in the blue regions. The initial state is $|0\rangle_L$. (a) Full connectivity in the device, and the modified decoder is the appropriate full-connectivity ZZ decoder. (b) Ring connectivity in the device, and the modified decoder is the appropriate ring decoder. (c) Ring connectivity in the device, and the modified decoder is the non-matching FC ZZ decoder.
The range of parameters for which the modified decoder is advantageous (red) significantly decreases when the modified decoder is not chosen to match the connectivity in the device.
}
\end{figure}

\section{A logical error with the Pauli crosstalk channel}\label{sec:PauliCounting}

In this subsection we point to the reason that the Pauli channel constructed from the 2Q crosstalk generators of the connected pairs could deviate from the unitary dynamics that it is supposed to describe, when the full connectivity ZZ decoder is applied, to some input states.

For this purpose we define the crosstalk coefficients
\begin{equation}
    c = \cos\frac{\zeta t}{2}, \qquad s= \sin\frac{\zeta t}{2}.
\end{equation}
The action of a Pauli noise channel composed of $n$ compositions of the 2Q Pauli channel defined in \eq{eq:P2Q} can be written in terms of a vectorized superoperator as $\hat{P} \dket{\rho}$, and $ \hat{P}$ expanded to second order terms as
\begin{multline}    
    \hat{P} = c^{2n}\hat{\identity} + c^{2(n-1)}s^2 \sum_{\langle i,j\rangle} \hat{Z}_i \hat{Z}_j + \\ c^{2(n-2)}s^4 \sum_{\langle i,j\rangle} \hat{Z}_i \hat{Z}_j \sum_{\langle k,l\rangle\neq \langle i,j\rangle} \hat{Z}_k \hat{Z}_l + ...
\end{multline}
where the summation is over the connected pairs and it is understood that the hat notation of each of the terms in the above expansion indicates that it acts simultaneously on the left and on the right of the density matrix (as discussed in \app{App:commuting_channels}).

The unitary operator constructed from the application of the same generators would be
\begin{equation}
    U=\exp\left\{-i\sum_{\langle i,j\rangle}\frac{\zeta t}{2}Z_i Z_j\right\} = \prod_{\langle i,j\rangle}\left(c\identity - isZ_i Z_j\right),
\end{equation}
which  can be expanded to the same order of terms,
\begin{equation}
    U= u_0 + u_1 + u_2 + ... 
\end{equation}
with
\begin{equation}
    u_0 = c^n\identity, \qquad u_1= (-i)c^{n-1} s \sum_{\langle i,j\rangle}Z_i Z_j, \ee
and
\begin{equation} 
    u_2 = (-1)c^{n-2}s^2 \sum_{\langle i,j\rangle} Z_i Z_j \sum_{\langle k,l\rangle\neq \langle i,j\rangle} Z_k Z_l.
\end{equation}

The action of $U$ is given by $\rho \to U\rho U^\dagger$, and we are interested in spotting the leading-order difference from the action of $\hat{P}$ that gives an error not correctable by the code. Since the code corrects any $ZZ$ error, the leading order difference would come from a combination of four different (non-cancelling) $Z$ factors, i.e., \begin{equation}
Z_iZ_jZ_kZ_l,\quad i\neq j\neq k \neq l, \label{eq:ZZZZ}   
\end{equation}
which is equivalent to the logical $Z_L$ up to the missing qubit's $Z_m$, and could result in $Z_L$ being applied to the state if the syndrome would be interpreted (and corrected) as the single-qubit error $Z_m$. When the input state is not an eigenvalue of the logical $Z_L$, this results in a logical error.

Terms as in \eq{eq:ZZZZ} appear in the expansion of $U\rho U^\dagger$ in leading order in the term $u_2 \rho u_2^\dagger$, which is explicitly,
\begin{equation} 
    c^{2(n-2)}s^4 \sum_{\langle i,j\rangle} Z_i Z_j \sum_{\langle k,l\rangle} Z_k Z_l(\rho) \sum_{\langle i',j'\rangle} Z_{i'} Z_{j'} \sum_{\langle k',l'\rangle} Z_{k'} Z_{l'},
    \label{eq:u2rhou2}
\end{equation}
with $\langle i,j\rangle\neq \langle k,l\rangle$ and $\langle i',j'\rangle\neq \langle k',l'\rangle$.
The above expression should be contrasted with the Pauli construction of $\hat P$ that gives at the same order of terms (with $\langle i,j\rangle\neq \langle k,l\rangle$ as before),
\begin{equation} 
    c^{2(n-2)}s^4 \sum_{\langle i,j\rangle} Z_i Z_j \sum_{\langle k,l\rangle} Z_k Z_l(\rho)  Z_{i} Z_{j} Z_{k} Z_{l},
    \label{eq:P2rho}
\end{equation}
and a simple combinatorial counting shows that the expression in \eq{eq:u2rhou2} contains three times more terms that do not cancel and lead to a logical error larger by a factor of three as compared with \eq{eq:P2rho}. This is easiest to see by fixing one term on the left of $\rho$ in \eq{eq:u2rhou2}, e.g., $(i,j,k,l)=(1,2,3,4)$, which can be matched by three terms to the right of $\rho$, i.e.,
\begin{equation}
    (i',j',k',l')=(1,2,3,4),\,(1,3,2,4),\,(1,4,2,3),
\end{equation}
of which only the first exists in the Pauli construction. This results in a factor of 3 that in this case -- due to the $s^4$ dependence if \eq{eq:P2rho} -- can be corrected for small $|\zeta|t$ by a scaling of the noise by $\zeta\to \sqrt[4]{3}\zeta$ to result in the same logical error (up to higher orders in $\zeta$).

The above expansion could perhaps be generalized numerically or circumvented with some scaling factors when building the Pauli noise, in order to accommodate for qubit-dependent phase terms, and other types of noise and qubit connectivities.

\section{Modified Decoder details}\label{sec:mod-dec-details}

The error correction procedure is composed of the following steps - encoding, decoding, and performing the recovery. 
We will focus on the decoding step where syndromes are mapped to
the recovery process required to restore a code state. A stabilizer
measurement resulting in $+1$ or $-1$ reflects as a $0$ or $1$
in the syndrome accordingly. In a logical code state, all stabilizer
measurements yield +1, resulting in a syndrome of 0000. Any deviation
from this syndrome indicates a departure from a logical state, necessitating
a recovery process.

The standard decoder associates each non-zero syndrome with a 1Q
Pauli error, assuming that 1Q Pauli errors are the most
prevalent. However, with the Lindbladian introduced in \seq{sec:lindbladian},
the likelihood of a $Z$ error on two connected qubits may be comparable
to, or even greater than, the probability of a 1Q Pauli
error, depending on the parameter values. The connectivity between qubits dictates the probability
of 2Q errors. As an illustration, we will examine two connectivity
scenarios: one where all qubits are interconnected and another where
each qubit $i$ is only connected to its neighboring qubits $i-1({\rm mod}\,5)$ and $i+1({\rm mod}\,5)$, forming a
ring structure. For each connectivity scenario, we define a decoder
tailored to its specific characteristics. 
The standard decoder can be slightly modified to gain the ability to correct $Z_iZ_j$ errors for $j=i+1({\rm mod}\,5)$ while sacrificing the ability to correct singe qubit $Y$ errors. we denote this decoder the ring decoder. 
Similarly, it can be modified to gain the ability to correct all $Z_iZ_j$ errors while sacrificing the ability to correct single qubit $X$ or $Y$ errors, which we denote as the full-connectivity ZZ (FC ZZ) decoder.
The details of these decoders
are presented in Table \ref{tab:mod-dec}.

For a broad spectrum of parameter ranges, as can be seen in \fig{fig:modified decoder} and \fig{fig:modified decoder all}, the decoder tailored to the connectivity
of the device demonstrates superior performance compared to the standard
decoder. 
When the decoder does not fit the connectivity in the device, the range of parameters where it is useful decreases significantly.
This highlights the significance of considering the specific
error model and the specific connectivity characteristics of the quantum
device when optimizing decoding strategies.

\bibliography{citations}

\end{document}